\documentclass{iopart}
\usepackage{iopams}
\usepackage{epsfig}
\usepackage{euscript}
\def \esc {\gamma_{\rm esc}}
\def\a{\alpha}
\def\b{\beta}
\def\g{\gamma}
\def\G {\Gamma}
\def\d{\delta}

\def \e{\varepsilon}

\def\Tr{\mbox{tr}\,}
\def\S{{\cal S}}
\def\l{\lambda}
\def\L{\openone}
\def\m{\mu}
\def\o{\omega}
\def\O{\Omega}
\def\DD{\partial}

\def\Bolts {k_{\rm B}}

\def\dwell{\tau_{\rm d}}
\def \Dif {{\cal D}}
\def\Coop {{\cal C}}

\def\openone{{\bf 1}}
\def\onlinecite{\cite}
\begin{document}
\title[Charge Flucutations]{Charge fluctuations in open chaotic cavities}
\author{M. B\"uttiker and M. L. Polianski}
\address{\ D\'epartement de Physique Th\'eorique, Universit\'e de Gen\`eve, CH-1211 Gen\`eve 4,
Switzerland}
\date{\today}

\begin{abstract}
We present a discussion of the charge response and the charge
fluctuations of mesoscopic chaotic cavities in terms of a
generalized Wigner-Smith matrix. The Wigner-Smith matrix is well
known in investigations of time-delay of quantum scattering. It is
expressed in terms of the scattering matrix and its derivatives with
energy. We consider a similar matrix but instead of an energy
derivative we investigate the derivative with regard to the electric
potential. The resulting matrix is then the operator of charge.  If
this charge operator is combined with a self-consistent treatment of
Coulomb interaction, the charge operator determines the capacitance
of the system, the non-dissipative ac-linear response, the RC-time
with a novel charge relaxation resistance, and in the presence of
transport a resistance that governs the displacement currents
induced into a nearby conductor. In particular these capacitances
and resistances determine the relaxation rate and dephasing rate of
a nearby qubit (a double quantum dot). We discuss the role of
screening of mesoscopic chaotic detectors. Coulomb interaction
effects in quantum pumping and in photon assisted electron-hole shot
noise are treated similarly. For the latter we present novel results
for chaotic cavities with non-ideal leads.
\end{abstract}
 \maketitle

\section{Introduction}\label{sec:Introduction}

Quantum transport in structures so small that the quantum wave
nature of particles becomes important presents many theoretical and
experimental challenges. The physical realization of such small
structures is quite diverse and ranges from electronic transport
through (partially) coherent samples
\cite{nato,Beenakker,Alhassid,review,ABG} to scattering of photons
in cavities \cite{Doron_90,cavity} or scattering of particles in
compound nuclei \cite{nucleus}. Often the dynamics of particles
inside the scattering region is chaotic. In an optical cavity this
is a consequence of the irregular shape of the resonator and,
similarly, in electrical samples chaotic scattering results from the
boundary or the impurity configuration. This paper, devoted to
electron transport, is focused mostly on the particular example of a
chaotic cavity, namely a two-dimensional chaotic quantum dot. A
typical quantum dot is usually patterned into the two-dimensional
electron gas at the interface of semi-conducting heterostructures.
The two-dimensional electron gas results from strong quantization of
the electron motion perpendicular to the interface. The shape of the
sample is formed with the help of metallic gates on top of the
structure. Even if the dot is free of impurities, the shape of the
dot is usually quite irregular, so that the dynamics of electrons is
classically chaotic. (If in addition there are impurities chaos
results independently of the geometric shape of the dot).

At sufficiently low temperatures, when the length over which the
carriers retain phase memory becomes comparable to the dimensions of
the structure we enter the regime of \emph{mesoscopic}
physics \cite{MPiS,LesHouches}. In this regime, every dot exhibits
fluctuating physical properties due to the high sensitivity of
quantum interference effects on the exact geometry of the dot. Then
properties of small structures must be described by their
\emph{statistics} in the form of mesoscopic (sample-to-sample)
distribution functions rather then by their averages over ensemble
of similarly fabricated samples.

We are interested in dots which are connected to the outside via one
or several leads which permit the exchange of carriers with
electrical contacts (reservoirs). The connection between the dot and
reservoir can be highly transparent or alternatively we can insert
tunnel barriers which increasingly insulate the dot from the
contacts. If the dot is closed, its equilibrium charge changes in
response to the voltage applied to the external gates. Except for
special gate voltages the charge on the dot does not fluctuate. For
poorly transmitting contacts (strong tunnel barriers) the charge of
the dot is strongly quantized and at low temperatures this
quantization can block transport (Coulomb blockade). Such a blockade
is important if a typical energy needed to add an electron into a
dot, $\sim e^2/C$ (capacitance $C$ defined by geometry), is large
compared to the temperature and the escape rate $\esc$ of carriers
from the dot, $e^2/C\gg \Bolts T,\hbar\esc$ \cite{MPiS}. However,
when the barrier becomes transparent and the contacts are wide open
the charge quantization vanishes and weak charge fluctuations appear
at all voltages.

The main subject of this paper is the role of Coulomb interactions
on transport properties of the dots which are connected with highly
transparent contacts to reservoirs. It turns out, that in many
transport problems the role of Coulomb interactions is closely
connected to the Wigner-Smith matrix \cite{wigner,smith}. This matrix
is well known in investigations of the time-delay of quantum
scattering \cite{Fyodorov_Sommers_96,gopar,Fyodorov_JMPh, Lehmann,bfb,waves}. It is expressed
in terms of the scattering matrix and its derivatives with energy.
We consider a similar matrix but instead of an energy derivative we
investigate the derivative with regard to the electric
potential \cite{mb82,mb83}. The resulting matrix is then the operator
of charge. If this expression for the charge operator is combined
with a description of interaction on the Hartree level (random phase
approximation) \cite{AA} the charge operator thus found plays an
important role in a number of transport problems. The list of these
problems includes charge rearrangement in weakly non-linear
transport, the current-response to oscillating potentials,
relaxation and dephasing in weakly coupled nearby conductors, the
theory of quantum detectors, adiabatic quantum pumping, and
frequency-dependent thermal and shot noise.

There are several important questions one might ask: What are the
signatures of Coulomb interactions in transport properties of a
sufficiently open quantum dot? Which experiment could probe the
Coulomb interactions? Can Coulomb effects be distinguished from
other effects like dephasing? Of course for poor contacts the
Coulomb blockade has a very clear signature in transport. However,
many experiments are performed with few- and multi-channel open
dots, so it is also important to account for Coulomb interactions in
such samples.

Our discussion is based on the scattering matrix approach to chaotic
systems \cite{BlumelSmilansky,Baranger_Mello,jal}. The formalism is
a subject of many extensive reviews
\cite{Beenakker,Alhassid,ABG,MPiS}, and we just outline here why
this method is advantageous for our purposes. Our starting point is
the energy-dependent scattering matrix $\S(\e)$ that relates
incoming and outgoing electrons at energy $\e$. This matrix is
obtained from the solution to the Schr\"odinger equation for
\emph{non-interacting} electrons. For low frequencies $\o\ll \esc$,
the scattering matrix varies weakly on the scale $\sim \o$, so for
linear transport one can expand transport coefficients in $\o$. It
is this expansion which makes the Wigner-Smith matrix appear. For
quantum dots the statistical properties of the Wigner-Smith matrix
have recently been widely explored
\cite{Fyodorov_Sommers_96,gopar,Fyodorov_JMPh,Lehmann,bfb,waves,Savin},
so that one can use them to find the statistical distribution of
measured quantities.

The scattering matrix approach is not the only possible method to
solve transport problems. For a discussion of alternative methods
and applicability of our approach we refer to Sec.
\ref{sec:discussion}. The self-consistent treatment of the Coulomb
interaction is the subject of our paper, and we apply this
method even for few-channel quantum dots.

We proceed as follows: In Section \ref{sec:WignerSmith} we first
consider non-interacting electrons and introduce the Wigner-Smith
matrix as a natural building block for a low-frequency
low-temperature transport theory of open cavities. On the example of
a single-channel cavity we demonstrate that taking into account the
Coulomb interactions is essential in finite frequency transport. A
self-consistent treatment allows us to consider the effect of
Coulomb interactions and find the internal potential and its
derivative with respect to external perturbations for a dot with
arbitrary number of channels. Later we concentrate on applications
of our approach to various experimental set-ups. Sec.
\ref{sec:CandR} considers the role of Coulomb interactions in the
renormalization of the effective capacitance and introduces the
mesoscopic charge relaxation resistance in terms of the Wigner-Smith
matrix. These results are used later in Sections
\ref{sec:shotnoise}--\ref{sec:acnoise}. Sec. \ref{sec:shotnoise}
considers the displacement current induced into a nearby gate due to
the charge fluctuations associated with shot-noise in a dc-biased
two-channel cavity and finds its mesoscopic distribution. In Sec.
\ref{sec:qubit} the relaxation rate and dephasing rate of a charge
qubit near a chaotic cavity are discussed. In Sec.
\ref{sec:detector} we describe how to use these results to describe
a quantum detector. Sec.  \ref{sec:pump} concerns 'quantum pumps',
which pump electrons due a time-periodic variation of the dots
shape. The effect of Coulomb interactions on the pumped voltage is
considered self-consistently. As another application of the Hartree
treatment we consider an ac-biased multi-channel quantum dot in the
Sec. \ref{sec:acnoise} and find the photon-assisted shot-noise both
for completely open dots and partially open dots with equal channel
transparencies. In this section we use results of \ref{sec:appendix}
for various correlators of the scattering matrices. We discuss other
approaches and the applicability of our results in Sec.
\ref{sec:discussion} and conclude in Sec. \ref{sec:conclusions}.


 \section{Generalized Wigner-Smith matrix}\label{sec:WignerSmith}

In this work we consider electrical transport problems which can be
described with the help of a generalized Wigner-Smith matrix
\cite{wigner,smith}. We illustrate these transport problems for
chaotic cavities. In linear response to oscillating potentials the
Wigner-Smith matrix describes the leading non-dissipative response
and in certain cases also the leading dissipative term. In
particular, capacitances \cite{btp}, charge relaxation resistances
\cite{btp}, and a Schottky resistance in the presence of shot noise
\cite{plb}, can be expressed in terms of interaction constants and
the Wigner-Smith matrix.

\begin{figure}
\epsfxsize=.7\hsize
\centerline{\epsffile{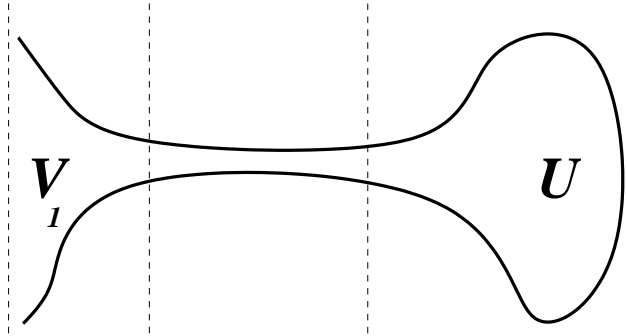}} \vspace*{0.3cm} \caption{
\label{cavity} Mesoscopic capacitor connected via a single lead to
an electron reservoir. $V_1$ is the voltage applied to the contact,
$U$ is the potential in the cavity.}
\end{figure}

To be specific, consider the structure in Fig. \ref{cavity}. A cavity is
connected with a single lead to a "contact". At the Fermi energy
there are $N$ open quantum channels. As a consequence there exists a
scattering matrix $\S$ of dimension $N \times N$ with elements
$\S_{mn}$ which relates the current amplitude of the incoming state
$n$ to the current amplitudes of the out-going state $m$. Each
carrier incident on the cavity will after some delay be reflected
and leave the cavity to be scattered back into the contact.
According to Wigner and Smith the time carriers with energy $E$
spend in the cavity can be found from the matrix,

\begin{equation}
{\cal N} = \frac{1}{2\pi i} \S^\dagger \frac{d\S}{dE} . \label{ws}
\end{equation}
Taking the trace and multiply with Planck's constant gives a sojourn
time $\tau_s = h\, \Tr {\bf \cal{N}}/N$. The time defined in this
way depends on how exactly we define the scattering matrix $\S$. We
could define $\S$ right at the entrance to the cavity
(right most broken line in Fig. \ref{cavity}), somewhere in
the long lead, at the entrance to the contact, or even deep inside
the contact (left most broken line in Fig. \ref{cavity}).
In the original work \cite{smith} the $\S$-matrix is
clearly defined in an asymptotic way on the surface of a sphere with
a radius that is eventually taken to infinity. The time calculated
by Smith is a delay, that is a difference of a time in the presence
and absence of a scattering center. This is a sensible procedure if
we are not interested exactly where a carrier incurs the delay.

Our aim, however, is to obtain an expression for the {\it charge},
not the time-delay. Moreover, we are interested in the local charge.
We have to know, whether this charge is accumulated inside the
cavity, in the lead, or inside the contact. Therefore, the
Wigner-Smith matrix is not the appropriate object \cite{mb82,mb83}.
The {\it local} question which we are asking can of course also be
asked for time. For instance, the widely discussed question, "How
much time does a carrier spend while traversing a tunnel barrier?"
is a question about local properties \cite{mb82,mb83} which can not
be answered by an appeal to the Wigner-Smith matrix. It was already
noticed in the treatment of this problem \cite{leavens}, that
instead of using derivatives with regard to energy to find delay, we
have to consider derivatives with regard to the electric potential
$U$ to find its conjugate, charge. Therefore, consider a small
change of the potential inside the cavity only. We can then consider
the matrix

\begin{equation}
{\cal N} = - \frac{1}{2\pi i}\S^\dagger \frac{d\S}{d(eU)} .
\label{bu1}
\end{equation}
Here the $\S$-matrix can be defined asymptotically, \emph{i.e.} in
the contact to the left of Fig. 1. Nevertheless, the answer given by
Eq. (\ref{bu1}) depends now only on the effect which a small change
of potential inside the cavity generates. We can now introduce the
dwell time \cite{mb83}, which is the total integrated density in the
cavity in an energy range $dE$ divided by the incoming carrier flux
$(N/h)dE$
\begin{equation}
\dwell= \frac{h}{N} \Tr {\bf \cal{N}} = i \frac{{\hbar}}{N}\Tr
\S^{\dagger} \frac{d\S}{d(eU)} . \label{bu2}
\end{equation}
We can choose any other potential, in the lead, or in the contact,
and ask in a similar way what the dwell time is in the region of
interest. We can take the region to be arbitrarily small and instead
of a derivative of the $\S$-matrix investigate the functional
derivative with regard to the local potential
\cite{mb93,mbam,mbam1}. For the purpose of this work, it is,
however, sufficient to consider derivatives with regard to single
potential $U$.

Instead of time-delay, or more precisely dwell time, we
alternatively use the language of density of states. The density of
states inside the cavity is given by

\begin{eqnarray}
\nu = \Tr {\cal N} = - \frac{1}{2\pi i} \Tr \left(\S^\dagger \frac{d
\S}{d(eU)}\right) = - \frac{1}{2\pi i} \frac{d}{d(eU)} \mbox{ln}
\left(\mbox{det }\S\right), \label{bu3}
\end{eqnarray}
where $U$ is the potential in the cavity. If $\nu$ is multiplied by
an energy interval $d E $ and the charge $e$, then $\Delta Q = e \nu
d E$ is the charge in the cavity due to the incident scattering
states in the interval $d E$. Importantly this is the charge of
non-interacting carriers. A conductor, when charged will respond
with an induced electrical potential $U$ to bring the charge in a
given region to the value permitted by the Coulomb interaction.

Suppose that we consider the linear response of the current in the
contact $dI(t) = dI(\omega) \exp(-i \omega t)$ to an oscillating
potential applied to the contact of the sample, $dV(t) = dV(\omega)
\exp(-i\omega t)$. In the zero-temperature limit the response is to
leading orders in frequency,

\begin{equation}
 G(\omega)  = dI(\omega)/dV(\omega) = - i \omega  e^{2}\Tr { \cal
N} + \omega^{2} (e^{2}  /2h)\Tr {\cal N }^2+ O(\omega^{3}).
\label{bu4}
\end{equation}
The first term is a non-dissipative response determined by the
density of states $\nu$ of the cavity. $e^{2} \nu$ has the the
dimension of capacitance and is called the quantum capacitance
\cite{btp,lamb,mb87,luryi}, $C_q = e^{2} \nu$. The second term is
real and thus represents the leading dissipative part of the
response.

Equation (\ref{bu4}) is not, however, a physically acceptable result
already for the geometry shown in Fig. 1. It is necessary to
consider the fact that electrons are interacting particles. To see
the problem, suppose that the cavity and/or the lead are charged
above what the ionic background of the conductor permits. As a
consequence there will be long range electrical field lines
emanating from the cavity and the lead. However, if the contact, the
lead, and the cavity are the only metallic bodies, the electrical
field lines emerging from the cavity or the lead will eventually end
on the metallic contact. There exists, therefore, a Gauss volume
taken large enough such that the electrical flux through its surface
vanishes. The charge inside this Gauss volume is a constant of
motion. Instead of Eq. (\ref{bu4}) we should find $G(\omega)  =0$.
We can see the profound effect of the long range Coulomb interaction
in another way. The particle current density $I^{\rm p}({\bf r},t)$
is not necessarily a conserved quantity. But if we add the Maxwell
displacement current $I^{\rm d}({\bf r},t) = \epsilon_L dE({\bf
r},t)/dt$ with $E({\bf r},t)$ the electric field and consider the
total current $I^{\rm tot} ({\bf r},t) = I^{\rm p}({\bf r},t) +
I^{\rm d}({\bf r},t)$ we have a quantity of vanishing divergence,
$\mbox{div }I^{\rm tot}({\bf r},t) = 0$. The total current has no
sources or sinks. It is important to emphasize that it is the total
current that is measured in experiment. Thus Eq. (\ref{bu4}), valid
for non-interacting carriers, is on physical grounds unacceptable.
\begin{figure}
\epsfxsize=0.7\hsize \centerline{\epsffile{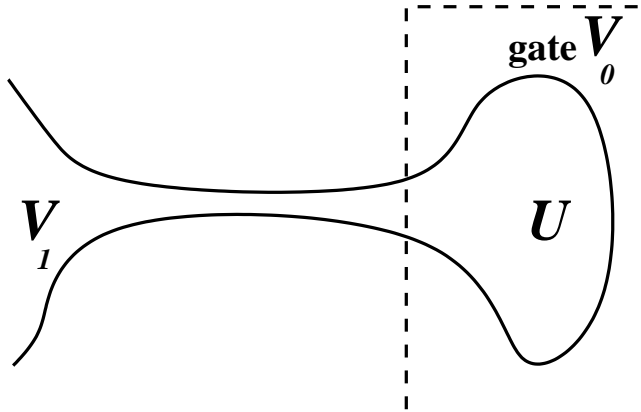}}
\vspace*{0.3cm} \caption{ \label{mesocap} Mesoscopic capacitor
connected via a single lead to an electron reservoir and
capacitively coupled to a gate. $V_1$ and $V_0$ are the potentials
applied to the contacts, $U$ is the electric potential of the
cavity.}
\end{figure}

We now consider a more general arrangement and explain how the
self-consistent Coulomb potential is determined. We consider
transport through a dot with several contacts at which voltages
$V_{\alpha}(t)=V_{\alpha}\cos(\o t+\phi_{\alpha})$ at some frequency
$\o$ are applied. In the end our results are also applicable for dc
transport, if the limit $\o\to 0$ is taken. In fact, in a dc-biased
system, within linear transport, the potential $U(t)=\mbox{const}$
and is not important, since at zero frequency, $\o=0$, the current
is gauge invariant and knowledge of the internal potential is not
necessary. However, it's derivative with respect to an external
perturbation in the $\a$-th lead, $dU/dV_\a$, can be used,
\emph{e.g.} in the analysis of the magnetic field asymmetry of the
current in the non-linear transport through a dot
\cite{DavidMarkus}. As  described above, the charge inflow into the
dot shifts the internal potential $U(t)$ due to capacitive coupling
to the gate $C$, kept at a potential $V_0(t)=V_0\cos(\o t+\phi_0)$.

The current $I_\a(\O)$ at finite frequency $\O$ in the $\a$-th lead
attached to the dot is the sum of two contributions: particle
current $I_\a^{\rm p}(\O)$ and displacement current $I_\a^{\rm
d}(\O)$. The former is due to variations of electro-chemical
potentials in various leads and is expressed via conductances
$G_{\a\b}(\O)$ \cite{btp} of non-interacting electrons
\begin{eqnarray}\label{eq:G}
\fl G_{\a\b}(\O)= \frac {e^2}{h}\int
d\e\,\Tr\left(\d_{\a\b}\openone_\a - \openone_\a{\cal
S}^\dagger(\e)\openone_\b{\cal
S}(\e+\hbar\O)\right)\frac{f(\e)-f(\e+\hbar\O)}{\hbar\O}.
\end{eqnarray}
Here we introduce the matrices $\L_\a$ which have unit elements
along the diagonal of the channels of contact $\alpha$ and zero
otherwise. The displacement current, the result of screening,
corresponds to variations of charge inside the dot and is expressed
through the potential $U(\O)$ and yet unknown susceptibility
$\chi_\a(\O)$. Thus the current at probe $\a$ is
\begin{eqnarray}
I_\a(\O)=I_\a^{\rm p}(\O)+I_\a^{\rm d}(\O)=\sum_\b
G_{\a\b}(\O)V_\b(\O)+\chi_\a(\O) U (\O).
\end{eqnarray}
We require gauge-invariance, that implies that the global shift of
all potentials $V_i(\O) \to V_i(\O)-f(\O), U(\O)\to U(\O)-f(\O)$ by
an arbitrary $f(\O)$ does not change the \emph{total} current
\cite{btp}. This determines the susceptibility,
$\chi_\a(\O)=-\sum_{\beta}G_{\alpha\beta}(\Omega)$. As a consequence
the current depends on difference of potentials only, as expected:
\begin{eqnarray}
I_\a(\O)&=& \sum_\b G_{\a\b}(\O)(V_\b(\O)-U(\O)). \label{eq:invar_I}
\end{eqnarray}
On the other hand, from charge conservation
$\sum_{\alpha}I_{\alpha}(t)=C(d/dt)[V_0(t)-U(t)]$, we find that
$U_\O\neq 0$ only at $\O=\pm\o$. In response to potentials with
Fourier components $V_{\a,\o}=V_\a \exp(i\phi_\a)$ the
frequency-dependent potential $U_\o$ is
\begin{eqnarray}\label{eq:UU}
U_\o= \frac{\sum_{\a\b}G_{\a\b}(\o)V_{\b,\o}+i\o
CV_{0,\o}}{\sum_{\a\b}G_{\a\b}(\o)+i\o C}.
\end{eqnarray}
At low frequencies, which is often important, the scattering
matrices $\S$ are weakly dependent on energy, and the first terms in
an expansion of the scattering matrix in energy are sufficient to
evaluate the potential $U_\o$ and the partial derivative of $U_\o$
with respect to the potential $V_{\a,\o}$. With the Wigner-Smith
matrix $\cal N$ introduced in Eq. (\ref{ws}) and the matrix of
voltages $\hat V_\o=\sum_\a\L_\a V_{\a,\o}$ the sample-specific
expressions read
\begin{eqnarray}\label{eq:U}
U_\o-V_{0,\o}= \frac{\Tr 
{\cal N}(\hat V_\o-\L V_{0,\o})}{C/e^2+\Tr 
{\cal N}},\,\,\,\frac{\DD U_\o}{\DD V_{\a,\o}}=\frac{\Tr {\cal
N}\L_\a}{C/e^2+\Tr {\cal N}}.
\end{eqnarray}
Both quantities naturally vanish in the non-interacting limit
$C\to\infty$ and reach a finite value in the opposite,
strongly-interacting limit, $C\to 0$. In the following we use this
approach to calculate several transport properties.


\section{Capacitance and charge relaxation resistance}\label{sec:CandR}
Fig. \ref{mesocap} shows a single-lead cavity which is separated by
an insulating material from a back gate at voltage $V_0(t)$. Now we
have the possibility to drive an ac-current through the system. In
addition to the current response to $V_1 (t)$ we can investigate the
response to $V_0 (t)$. A consistent treatment of the long range
Coulomb interaction demands that these two responses are different
only in sign. In fact the response must depend only on the voltage
difference $V(t) = V_1 (t) - V_0 (t)$, see Eq. (\ref{eq:U}). Here we
describe the Coulomb interaction in simple terms using a geometrical
capacitance $C$ which links the potential difference between cavity
and gate and the charge on the cavity and the gate. As a consequence
the conductance of the cavity capacitively coupled to a gate is now
given by
\begin{equation}
G(\omega) =  -i \omega C_{\mu} +  \omega^{2} C_{\mu}^{2} R_{q}  +
O(\omega^{3}). \label{gint}
\end{equation}
Here the first term is the electrochemical capacitance \cite{btp}
which is the series capacitance of the geometrical capacitance $C$
and the "quantum capacitance" $\nu e^{2}$,
\begin{equation}
C_{\mu}^{-1} = C^{-1} + (\nu e^{2})^{-1} . \label{cmu}
\end{equation}
Thus the capacitance is not a purely geometrical quantity, but
depends via the density of states $\nu$ on the properties of the
specific electrical conductor. Notice, that $C_\m^{-1}$ is a
fluctuating quantity, with mesoscopic average equal to $\langle
1/C_\m\rangle =1/C+\Delta/e^2$. Here $\Delta$ is the mean level
spacing in the dot. Equation (\ref{cmu}) has important implications:
for instance, if the cavity is deformed into a ring with an
Aharonov-Bohm flux through the hole of the ring, the capacitance
exhibits quantum oscillations which are periodic in the flux
\cite{physica,physica1}. Such oscillations have been observed in
arrays of rings by Deblok \emph{et al.} \cite{deblock} and in a
strikingly clear manner on a single mesoscopic capacitor by Gabelli
{\it et al.} \cite{gabelli}.

The second, dissipative term, is governed by a resistance $R_q$
which we call a charge relaxation resistance \cite{btp}:
\begin{equation}
R_{q} = \frac{h}{2e^{2}}\frac{\Tr {\cal N}^2 } {\Tr^2{\cal N}} .
\label{rq}
\end{equation}
Equations (\ref{cmu},\ref{rq}) are given in the low temperature
limit $\Bolts T\ll\hbar \o$. To understand the meaning of these
quantities better it is instructive to consider a basis in which the
scattering matrix is diagonal. All eigenvalues of the scattering
matrix are of the form $\exp(i\zeta_{n})$ where $\zeta_{n}$ is the
phase which a carrier accumulates from multiple scattering inside
the cavity. Thus the density of states (\ref{bu3}) can also be
expressed as
\begin{equation}
\nu = - (1/2\pi) \sum_{n}(d\zeta_{n}/d(eU)) \label{denp} .
\end{equation}
Similarly we can express the charge relaxation resistance in terms
of the potential derivatives of phases. In the low temperature limit
$R_q$ is determined by the sum of the squares of the dwell times
divided by the square of the sum of the dwell times,
\begin{equation}
R_{q} = \frac{h}{2e^{2}}\frac{\sum_{n} (d\zeta_n/d(eU))^{2}}{(\sum_n
d\zeta_{n}/d(eU))^{2}}. \label{rqp}
\end{equation}
We now briefly discuss the charge relaxation resistance. First we
note that the resistance unit is not the resistance quantum
$h/e^{2}$ but $h/2e^{2}$. The factor two arises since the cavity is
coupled to one reservoir only. Thus only half the energy is
dissipated as compared to dc-transport through a two terminal
conductor. We emphasize that the factor two is not connected to
spin:  $h/2e^{2}$ results from a single spin polarized channel. In
the single channel limit, Eq. (\ref{rqp}) is {\em universal} and
given just by $h/2e^{2}$. This is astonishing since if we imagine
that a barrier is inserted into the lead connecting the cavity to
the reservoir one would expect a charge relaxation resistance that
increases as the transparency of the barrier is lowered. Indeed, if
there is a barrier with transmission probability $\cal T$ per
channel in the lead connecting the cavity and the reservoir, then in
the large channel limit, for ${\cal T}N \gg 1$, $R_q$ is
\begin{equation}
R_q = (h/e^{2}) (1/{\cal T}N). \label{bee}
\end{equation}
In the large-channel limit, Eq. (\ref{bee}) is inversely
proportional to the total transmission $N{\cal T}$. Thus in the
large channel limit Eq. (\ref{rqp}) behaves as expected. For a
one-channel connection to a reservoir the recent experiment by
Gabelli {\it et al.} \cite{gabelli} indeed finds a resistance of 12
k$\O$ for a mesoscopic capacitor formed with the help of a quantum
point contact in high magnetic fields in the quantum Hall regime. A
brief overview of charge relaxation resistances in mesoscopic
systems is presented in Ref. \cite{korea}.

\begin{figure}
\epsfxsize=.7\hsize \centerline{\epsffile{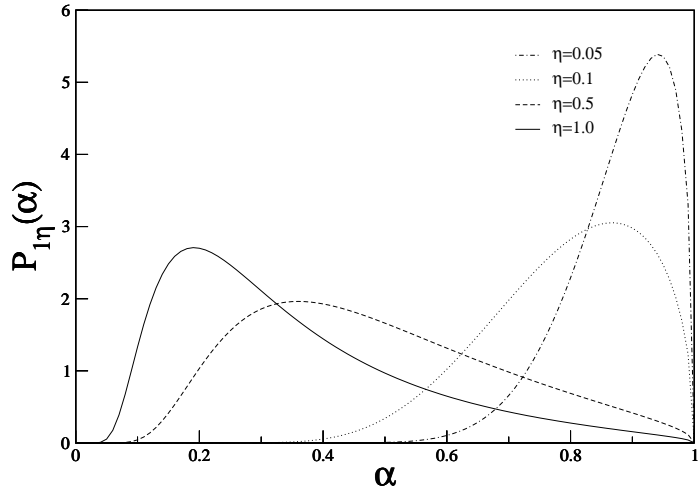}} \vspace*{0.3cm}
\caption{ \label{cap} Mesoscopic capacitance distribution
$P_{1\eta}(\a)$ of a cavity connected via a single perfect lead to a
reservoir. The capacitance is in units of the geometric capacitance,
$\alpha = C_{\mu}/C$. The ratio $\eta = \Delta (C/e^{2})$ is the
level spacing in units of the Coulomb energy. After Ref.
\protect\cite{gopar}. }
\end{figure}

For a single channel connected to a cavity, the charge relaxation
resistance is universal and given by  $h/2e^{2}$. Thus it is only
the capacitance $C_\m$ which needs to be investigated further.
Gopar, Mello and one of the authors \cite{gopar}  and independently
Fyodorov and Sommers \cite{Fyodorov_Sommers_96} calculated the
distribution of dwell times $w_{\beta}(\dwell )$ for a cavity
coupled to a lead with a perfect one-channel quantum contact in the
cases of time-reversal, broken time-reversal and broken
spin-inversion symmetry (denoted by Dyson symmetry indices $\beta =
1, 2, 4$ respectively \cite{Dyson}). This permits to find the
distribution function of the capacitance (\ref{cmu}) of such a
cavity. Fig. \ref{cap} shows the distribution $ P_{1 \eta}(\alpha)$
in the presence of time-reversal symmetry, $\b=1$, as a function of
$\alpha = C_{\mu}/C = \dwell /(\dwell + \eta ) $, where $\eta =
C\Delta/e^{2}$ is the ratio of the level spacing and Coulomb energy.
The distributions for larger $\eta$ are wider, because with
$\eta\to\infty$ the mesoscopic fluctuations of $\nu$ have a stronger
effect on $C_\m$. Typically the Coulomb energy is much larger than
the level spacing, the parameter $\eta$ is small and the
capacitances are close to the geometrical value (see Fig.
\ref{cap}).

\begin{figure}
\epsfxsize=0.7\hsize\centerline{\epsffile{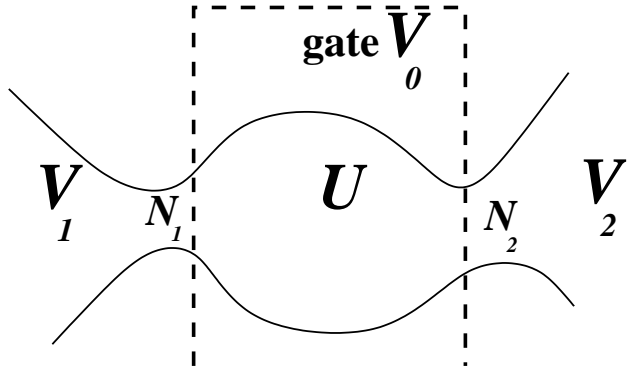}}
\vspace*{0.3cm} \caption{ \label{twolead} Chaotic dot connected via
two quantum point contacts with $N_1$ and $N_2$ open quantum
channels. The dot is capacitively coupled to a gate at voltage
$V_0$.}
\end{figure}

\begin{figure}
\epsfxsize=.5\hsize \centerline{\epsffile{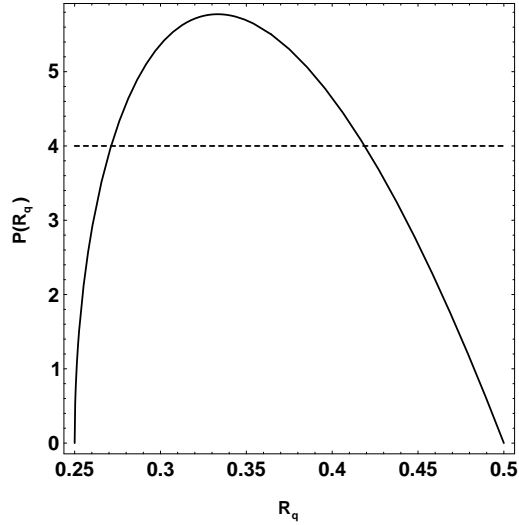}} \vspace*{0.3cm}
\caption{ \label{charger} Charge relaxation resistance $R_q$,
measured in units of $h/e^2$, for $\b=1$ (dashed curve) and $\b=2$
(solid). After Ref. \cite{plb}}
\end{figure}

Next let us consider a cavity that is connected via $N =2$ open
quantum channels to the outside. The two open channels can be
provided by the same contact (a single lead as in Fig.
\ref{mesocap}) or from different contacts as shown in Fig.
\ref{twolead}. There are now two dwell times which in general differ
from one another. Their joint distribution (for arbitrary $N$) is
derived in Ref. \cite{waves}. As a consequence we now have a
nontrivial distribution also of the charge relaxation resistance
\cite{plb} shown in Fig. \ref{charger}. For the case of
time-reversal symmetry, $\b=1$, Ref. \cite{plb} finds that $R_q$ is
uniformly distributed between $R_q = h/4e^{2}$ and $R_q = h/2e^{2}$,
whereas for the broken time-reversal symmetry, $\b=2$, the
distribution exhibits a peak and reaches zero at the limiting values
\cite{plb}. The limiting values are reached if both channels have
dwell times close to each other ($R_q = h/4e^{2}$) and if one of the
dwell times becomes very much longer than the other one ($R_q =
h/2e^{2}$). The probability of such processes for $\b=1$ is higher
then at $\b=2$, so that the distributions are stronger at
$R_q=h/2e^2,h/4e^2$.

\section{Charge noise in the presence of shot
noise}\label{sec:shotnoise}

We have derived the capacitance $C_{\mu}$ and the charge relaxation
resistance $R_q$ from linear response. The dissipative part of the
response proportional to $R_q$ is related to the charge and current
fluctuations of the system via the fluctuation dissipation-theorem.
Returning to the single-lead configuration in Fig. \ref{mesocap}, we
can calculate the current fluctuations at contact $\alpha = 1$ or at
the gate $\alpha = 0$ or their correlations, $S_{\alpha \beta} (t) =
1/2\int dt^{\prime} \langle\langle I_{\alpha}(t + t^{\prime})
I_{\beta}(t) + I_{\beta}(t) I_{\alpha}(t + t^{\prime})
\rangle\rangle$. Due to current conservation we have  $S(t) \equiv
S_{11} (t) = S_{00} (t) = - S_{10} (t) = -S_{01} (t)$. For the
Fourier transform of this noise spectrum we have $S_{II}^{\rm eq}
(\omega) = 2 \Bolts T \omega^{2} C_{\mu}^{2} R_q$, where we have
taken the classical limit $\hbar \omega \ll\Bolts T$. Since the
current is the time derivative of the charge accumulated on the
cavity we immediately also find the spectrum of the charge
fluctuations $Q$ on the cavity, $S_{QQ}^{\rm eq}(\omega) =  2 \Bolts
T C_{\mu}^{2} R_q$.

Next consider the case of a cavity with two contacts which permit
exchange of carriers with reservoirs, as shown in Fig.
\ref{twolead}. There $N_1$ quantum channels in the QPC connect the
cavity to the left contact at voltage $V_1$ and $N_2$ quantum
channels connect the cavity to the right at contact $2$ with voltage
$V_2$. If these two voltages differ, a transport state is
established. At zero temperature we have a current $I = G V$ with $V
= V_1 - V_2 $ determined by the Landauer formula, $G =
({e^{2}}/h){\cal T} $ where ${\cal T} = \Tr \S^{\dagger}\L_2
\S\L_1$. For a chaotic cavity the ensemble averaged total
transmission probability is ${\cal T} = N_1 N_2 /(N_1 + N_2)$. For
simplicity, we consider a voltage $eV \gg \Bolts T $ so large that
the thermal noise in this conductor can be neglected. However, the
granularity of charge and the fact that to every incident channel
there are different final channels leads to shot noise \cite{review}
with a zero-frequency noise spectrum given by $S_{II}^{\rm shot} =
2e ({e^{2}}/h) |V| \Tr(\L_1\S^{\dagger}\L_1\S\L_1 \S^{\dagger}\L_2
\S)$. In terms of the transmission eigenvalues the shot noise is
$S_{II}^{\rm shot} = 2e ({e^{2}}/h) |V| \sum_{n} T_{n} (1-T_{n})$.
For a chaotic cavity with large perfect quantum point contacts
$N_1,N_2\gg 1 $ the ensemble averaged shot noise
\cite{review,jal,Oberholzer} is $S_{II}^{\rm shot} = 2e ({e^{2}}/h)
|V| N_{1}^{2} N_{2}^{2} /(N_1 + N_2 )^{3}$. For single channel leads
$N_{1} = N_{2} = 1$, the case of interest here, shot noise is
characterized by a distribution which is given in Ref. \cite{plb}.
The stochastic transfer of carriers through the cavity leads to
fluctuations of the charge in the cavity as a function of time. Such
charge fluctuations can build up only if screening in the conductor
is not perfect. Therefore, consider now again a nearby gate which is
capacitively coupled to the conductor with a geometric capacitance
$C$. The fluctuating charge in the cavity induces charge
fluctuations in the nearby gate such that the charge on the
conductor and the gate is conserved. In the transport situation
considered here, the different voltages in the leads break the
symmetry between contacts. An important role is now played by
certain selected elements of the full Wigner-Smith matrix. The
elements which retain information on the contacts from which
carriers are incident are
\begin{equation} \label{qop2}
{\bf {\cal N}}_{\beta\gamma} (E) = - \frac{1}{2\pi i}\left(
\S^{\dagger}(E) \frac{d\S(E)}{d(eU)} \right)_{\b\g}.
\end{equation}
Introducing the second quantization operators ${\hat a}_{\beta}(E)$
which annihilate an electron in the incoming channel in lead
$\beta$, we write the charge operator \cite{plb} in the absence of
interaction as
\begin{equation}
e {\hat {\cal N}} = e {\bf {\cal N}}_{\beta\gamma} (E) {\hat
a}^{\dagger}_{\beta}(E) {\hat a}_{\gamma}(E).
\end{equation}

\begin{figure}
\epsfxsize=0.5\hsize\centerline{\epsffile{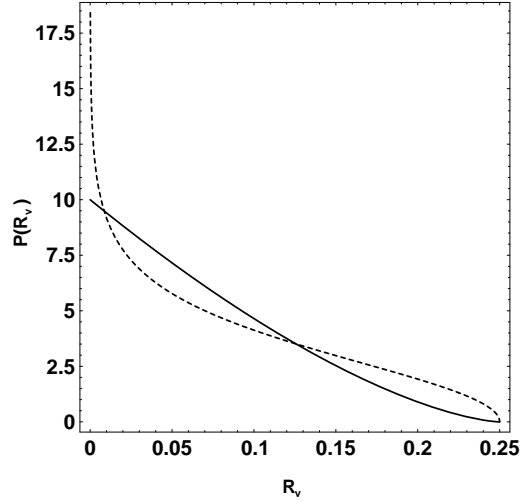}} \vspace*{0.3cm}
\caption{ \label{rv} Distribution of the resistance $R_v$ which
determines the low-frequency charge fluctuation spectral density of
a chaotic cavity connected to reservoirs with two single channel
leads. Dashed line is for the orthogonal ensemble, $\b=1$, solid
line is for the unitary ensemble, $\b=2$. After Ref. \cite{plb}}
\end{figure}

In reality the Coulomb interaction leads to potential fluctuations
inside the cavity and these fluctuations contribute in turn to the
charge. Since charge and potential are linearly related we are lead
to describe the fluctuating potential also with an operator ${\hat
U}$ that contributes to the net charge in proportion to the density
of states $\nu$ in the cavity. We have ${\hat Q}_{\beta\gamma} = e
{\hat {\cal N}}_{\beta\gamma} - e^{2}\nu {\hat U}_{\beta\gamma}$. On
the other hand the Coulomb interaction dictates that ${\hat
Q}_{\beta\gamma} = C {\hat U}_{\beta\gamma}$. Taken together, we
have therefore,
\begin{equation}
{\hat Q}_{\beta\gamma} = C {\hat U}_{\beta\gamma} = e {\hat {\cal
N}}_{\beta\gamma} - e^{2} \nu{\hat U}_{\beta\gamma}. \label{qop3}
\end{equation}
Solving for ${\hat U}_{\beta\gamma}$ determines the charge
fluctuations in the presence of screening,
\begin{equation}
{\hat Q}_{\beta\gamma} = \frac{C e {\hat {\cal
N}}_{\beta\gamma}}{(C+\nu e^{2})} = e \frac{C_{\mu}}{\nu e^{2}}
{\hat {\cal N}}_{\beta\gamma} . \label{dip}
\end{equation}
Here $C_{\mu}$ is the electrochemical capacitance, Eq. (\ref{cmu}).
Evaluation of the quantum statistical expectation value leads to a
low-frequency charge-fluctuation spectrum $S_{QQ}(0) = 2e|V|
C_{\mu}^{2} R_v$ determined by the resistance \cite{plb}
\begin{eqnarray}
R_v  = \frac{h}{e^2} \frac{\Tr
    {\cal N}\L_1 {\cal N}\L_2}
    {\Tr^{2}{\cal N}}.
    \label{Rv}
\end{eqnarray}
For a chaotic cavity connected only to two single channel leads the
distribution of the resistance $R_v$ is shown in Fig. \ref{rv}. In
many cases the mesoscopic distribution $P(C_\m)$ of the
electro-chemical capacitance $C_\m$ is much narrower then that of
$R_v$, cf. Fig. \ref{cap} for small $\eta$ and Fig. \ref{rv}, and
the fluctuations of noise $S_{QQ}$ are mostly due to fluctuations of
$R_v$.

It is interesting to calculate not only the spectrum of the noise
induced into the gate but also the spectrum at the contacts which
connect the cavity to reservoirs. Extending the discussion of Ref.
\cite{plb} Hekking and Pekola \cite{hp} find the finite frequency
noise spectra in the quantum limit. A quasi-classical discussion of
the frequency-dependent third cumulant of chaotic cavities is given
by Nagaev, Pilgram, and one of the authors \cite{npb}. It turns out
that in higher cumulants both the $RC$-time and the dwell time are
relevant \cite{npb}. For a related discussion of dynamical thermal
effects we refer the reader to Reulet and Prober \cite{rp}.


\section{Relaxation and dephasing rate of a charge
qubit}\label{sec:qubit}

\begin{figure}
\epsfxsize=.5\hsize \centerline{\epsffile{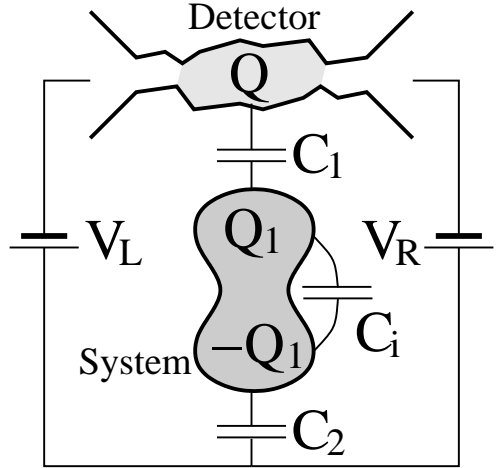}}
\vspace*{0.3cm} \caption{ \label{pilgram1} Chaotic cavity
capacitively coupled to a double quantum dot. After \cite{pilgram}}
\end{figure}

To illustrate the physical significance of the charge relaxation
resistance (\ref{rq}) and the resistance $R_v$ (\ref{Rv}), we now
consider a chaotic cavity coupled capacitively to a double quantum
dot as shown in Fig. \ref{pilgram1}. The double dot represents a
charge qubit: a single charge tunnels between the upper and lower
dot. The chaotic cavity plays the role of a detector. With this
set-up we can investigate the question \cite{pilgram}: if we use a
generic mesoscopic conductor (here a chaotic cavity) how good a
detector would this represent? The properties and suitability of
special systems, like quantum point contacts or single electron
transistors, as detectors have been widely discussed
\cite{mbam,pilgram,buks,hayashi,petta,gurvitz,aleiner1,levinson,Schoen1,Averin1,clerk}
and continue to be a subject of research
\cite{kicked,kicked1,Hartmann,counting,clerk2}. Interestingly a
chaotic cavity connected to one channel leads is with a probability
of almost one-half close to an ideal detector and if it is not ideal
changing the shape of the cavity just a bit leads with probability
of one-half to an almost ideal detector \cite{pilgram}.

The two-level system is represented by the Hamiltonian $\hat{H}_{DD}
= \frac{\epsilon}{2}\hat{\sigma}_z
 + \frac{\Delta}{2} \hat{\sigma}_x$, where
$\hat{\sigma}_i$ denote Pauli matrices. The energy difference
between upper and lower dot is $\epsilon$ and $\Delta$ accounts for
tunneling between the dots. The full level splitting is thus
$\Omega=\sqrt{\epsilon^2+\Delta^2}$. The chaotic cavity and the
two-level system are coupled through the long range Coulomb
interaction taken into account with a set of capacitances $C_1 ,C_2,
C_i $ that link the charge $Q_1$ and $-Q_1$ on the double dot to the
charge on the cavity. The series capacitance is $C^{-1} = C^{-1}_1 +
C^{-1}_2 + C^{-1}_i $.

If the two-level system is in a superposition state, it experiences
relaxation with a rate $\Gamma_{rel}$ towards the ground state due
to charge fluctuations in the cavity. In addition a coherent state
of the two level system is dephased with a rate $\Gamma_{dec}$.

At equilibrium the relaxation rate and decoherence rate generated by
the cavity are determined by the charge relaxation resistance
\cite{pilgram}
\begin{eqnarray}
\label{Central Result Relaxation} \Gamma_{rel} &= & 2\pi
\frac{\Delta^2}{\Omega^2} \left(\frac{C_{\mu}}{C_i}\right)^2
R_q \frac{\Omega}{2}\coth \frac{\Omega}{2\Bolts T},\\
\Gamma_{dec} &= & \frac{\Gamma_{rel}}{2} +
 2\pi \frac{\epsilon^2}{\Omega^2}
\left(\frac{C_{\mu}}{C_i}\right)^2 R_q \Bolts T .
\end{eqnarray}
A charge relaxation resistance exists even if the conductor permits
no transmission of carriers. Often the interaction of a qubit and a
detector is modeled by describing the effect of the qubit solely in
terms of a modulation of the transmission amplitude of the
conductor. Such a description predicts that a conductor with zero
transmission probability causes no relaxation and no dephasing. But
a conductor with transmission probability zero is just a capacitor,
and, as we have seen, a mesoscopic capacitor has a non-vanishing
charge relaxation resistance.

In the zero-temperature limit, the two-level system relaxes due to
the zero point fluctuations of the conductor. A qubit in an excited
state is a detector of vacuum fluctuations. If a voltage is applied
to the conductor, the decoherence rate contains in addition a term
which arises due to the charge fluctuations in the presence of shot
noise of the detector. This additional term is proportional to the
applied voltage and is governed by $R_v$,
\begin{equation}
\label{Decoherence RV} \Gamma_{dec} = \frac{\Gamma_{rel}}{2} +
 2\pi \frac{\epsilon^2}{\Omega^2}
\left(\frac{C_{\mu}}{C_i}\right)^2 R_v e|V| .
\end{equation}
Thus the density of states of the cavity $\nu$, its electrochemical
capacitance ${C_{\mu}}$, and the two resistances $R_q$ and $R_v$ are
the properties of the mesoscopic conductor which determine the
relaxation and decoherence rate of the two-level system.


\section{Chaotic cavity as a Quantum Detector}\label{sec:detector}

The cavity in Fig.  \ref{pilgram1} can be viewed as a detector of
the state of the two-level system. The current through the cavity
depends on whether the charge $Q_1$ is close to the cavity or in the
dot further away from the cavity. The difference of the currents
$\Delta I = I_1-I_2$ is evaluated using the Landauer formula
\begin{equation}
\Delta I = \Delta G |V| = \frac{e^2}{2\pi} |V| \sum
\frac{dT_n}{d(eU)} (e\Delta U)
\end{equation}
where $\Delta G$ is the change of conductance between the two states
of the double dot and $\Delta U = eC_{\mu}/((\nu e^2)(C_i-C_{\mu}))$
is the potential change on the mesoscopic conductor \cite{pilgram}.
To measure the current, the shot noise of the cavity must be
overcome. The zero-frequency shot noise is $S_{II} = e|V|(e^2/2\pi)
\sum T_n(1-T_n) $. Therefore, a measurement time
\cite{aleiner1,Averin1} $\tau_m = 4 S_{II} / (\Delta I)^2$ is needed
for a signal to noise ratio of 1. Using weak coupling $C_1,C_2 \ll
C_i$ one gets for the inverse measurement time \cite{pilgram}
\begin{equation}
\tau_m^{-1} = \Gamma_m = 2\pi \left(\frac{C_{\mu}}{C_i}\right)^2 R_m
e|V|
\end{equation}
with the resistance
\begin{equation}
R_m = \frac{h}{4e^{2}} \frac{(\sum_n dT_n/d(eU))^2} {(\sum_n
d\phi_n/d(eU))^2
 \sum_n T_n(1-T_n)  }.
\end{equation}
Fundamentally the measurement is always slower than the decoherence,
the decay of the off-diagonal elements of the reduced density matrix
of the two-level dot. This implies the inequality $\Gamma_m \le
\Gamma_{dec}$ or $R_m \le R_v$. This can be shown by deriving a
Schwarz inequality for a combination of scattering matrices and the
element ${\cal N}_{21}$ of the Wigner-Smith matrix (see Ref.
\cite{pilgram}) or more generally within linear response detector
theory \cite{clerk}.

An efficient detector provides a maximum of information for a
minimum of back-action (\emph{i.e.} decoherence) on the measured
system \cite{clerk}. The efficiency of the detector is determined by
the ratio $\eta = \Gamma_m /\Gamma_{dec} = R_m/R_v\leq 1$. The most
efficient measurement requires that the tunneling between the two
double dots is negligible, $\Delta \simeq 0$, and the temperature
must be much smaller than the applied voltage $\Bolts T \ll e|V|$.
But more importantly with the general formulation given above, we
can now discuss the conditions on the scattering matrix for the
detector to be ideal. Ref. \cite{pilgram} finds that the scattering
matrix needs to be of block-diagonal form: channel mixing detectors
are not ideal. Consequently the scattering matrix of an ideal
detector can be divided into $2\times 2$ blocks of the form
\begin{equation}
s^{(n)} = \left(
\begin{array}{cc}
-i\sqrt{1-T_n}e^{i(\phi_n+\phi_{A,n})} &
\sqrt{T_n}e^{i(\phi_n-\phi_{B,n})}\\
\sqrt{T_n}e^{i(\phi_n+\phi_{B,n})} &
-i\sqrt{1-T_n}e^{i(\phi_n-\phi_{A,n})}\\
\end{array}
\right).
\end{equation}
Each block is defined by its transmission probability $T_n$ and
three scattering phases $\phi_n$,$\phi_{A,n}$,$\phi_{B,n}$. Using
the definition of $R_v$ (Eq. (\ref{Rv}) we arrive at \cite{plb}
\begin{equation}
\label{Polar Decomposition} \fl R_v = \frac{h}{e^{2}} \frac{\sum_n
\left( (dT_n/dU)^2/(4 T_n (1-T_n)) + T_n (1-T_n)
(d(\phi_{A,n}-\phi_{B,n})/dU)^2 \right)} {\left(\sum_n
d\phi_n/dU\right)^2}.
\end{equation}
Eq. (\ref{Polar Decomposition}) reduces to earlier results
\cite{buks,aleiner1,levinson} in the infinite capacitance limit
where $C^{2}_{\mu} R_v$ in Eq. (4) can be replaced by $R_v(\nu
e^2)^2 $. In Eq. (\ref{Polar Decomposition}) the derivatives
$dT_n/d(eU)$ determine the sensitivity of the cavity to a potential
variation $\Delta U$. A high sensitivity implies a fast detector (a
short measurement time). However, a high sensitivity also implies a
fast decoherence rate proportional to $R_v$. From this we see that
decoherence and measurement speed are closely related quantities.

Demanding that $R_m$ and $R_v$ are
equal, requires ${d\phi_{A,n}}/{dU}- {d\phi_{B,n}}/{dU} = 0$. As
pointed out by Clerk, Girvin and Stone  \cite{clerk} , this condition is not
connected to any physical symmetry of the system, but can be
understood as a condition that the detector should not transfer
information into the phases of the scattered electrons since these
are not measured. However, we can require that each individual
phase-dependent term vanishes separately. $d\phi_{B,n} / d(eU) = 0$
in Eq. (\ref{Polar Decomposition}) can be achieved by requiring that
the scattering Hamiltonian must obey time-reversal symmetry. The
phases $d\phi_{A,n} / d(eU) = 0$ vanish for detectors that obey a
spatial inversion symmetry $V(x,y,z)=V(x,y,-z)$. (Here we assume
that conduction is along $z$ and $x,y$ are transverse coordinates).
Interestingly Ref. \cite{pilgram} finds that in the multichannel
case $N > 1$ another condition is
needed! The equality $R_m = R_v$ gives us a condition which is of
statistical origin. The total conductance of the detector is a sum
of one channel conductances that have independent uncertainties. The
statistical uncertainty of this sum is minimized if \cite{pilgram}
\begin{equation}
\label{Multichannel Condition} \frac{dT_n }{T_n(1-T_n) } = C(U)
d(eU).
\end{equation}
with a function $C(U) > 0$ that does not depend on the index $n$. In
the WKB limit we have $d/dE = -\partial/\partial(eU)$ and Eq.
(\ref{Multichannel Condition}) can be interpreted as differential
equations for the transmission probabilities $T_n (E)$. The
solutions are all of the form $T_n =
(1+e^{-\left(F(E)-F(E_n)\right)})^{-1}$ with $dF/dE = C$ (The
function $F$ is therefore monotonously increasing). The only
difference allowed between the different probabilities $T_n$ is the
offset energy $E_n$. Transmission probabilities of the type
(\ref{Multichannel Condition}) occur automatically if the scattering
problem is separable due to a potential of shape
\begin{equation}
\label{Transmission Condition} V(x,y,z) = Z(z) + W(x,y).
\end{equation}
This occurs in particular for the case $F = 2\pi E/ \omega_z$ with a
symmetric harmonic scattering potential $Z(z) = V_0 - m\omega_z^2
z^2 /2$ which is the saddle point potential of an ideal quantum
point contact.

To illustrate the role of the condition Eq. (\ref{Transmission
Condition}) we now consider chaotic cavities as detectors. The
condition Eq. (\ref{Transmission Condition}) states that a geometry
with a separable potential $V(x,y,z) = Z(z) + Y(x,y)$ is favorable
to obtain an efficient detector in the case of more than one open
channel. It is clear that a chaotic cavity violates this condition.
The efficiency $\eta=\Gamma_m / \Gamma_v = R_m / R_v$ of a chaotic
detector with two open channels in each contact is expected to be
much smaller than a chaotic cavity connected to single channel
contacts for which Eq. (\ref{Transmission Condition}) plays no role.

\begin{figure}
\begin{center}
\leavevmode \psfig{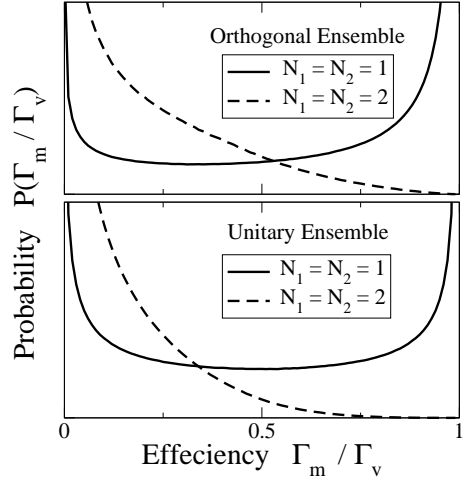}
\caption{Efficiency distribution of an ensemble of chaotic quantum
cavity detectors: orthogonal ensemble (top panel), unitary ensemble
(lower panel) for single channel ($N_1 = N_2 = 1$) and double
channel ($N_1 = N_2 = 2$) point contacts. After \cite{pilgram}}
\label{Chaos Plot}
\end{center}
\end{figure}

Using the distribution of the elements of the Wigner-Smith matrix
(the Laguerre ensemble \cite{bfb}) we find the probability
distribution of the measurement efficiency $\eta=R_m/R_v$ in the
orthogonal (time-reversal symmetry) and unitary ensemble (broken
time-reversal symmetry) shown in Fig. \ref{Chaos Plot}. The
distribution for $N_1=N_2=1$ in the unitary ensemble can also be
calculated analytically, $P(\eta)=1/\sqrt{\eta(1-\eta)}$. The
distributions for the other cases were obtained by numerical
integration. Interestingly Fig. \ref{Chaos Plot} shows that despite
the absence of inversion symmetry a chaotic dot with open single
channel contacts is with high probability an efficient detector! It
is clearly visible that chaos reduces strongly the efficiency of the
measurement device as soon as more than one channel contributes to
the electric transport. Compared ti the time-reversal symmetric case
(solid curve in the upper panel) the reduction due to a broken
time-reversal symmetry (solid curve in the bottom panel) is much
less pronounced.

\section{Quantum pumping}\label{sec:pump}

\begin{figure}
\epsfxsize=1.\hsize
\epsffile{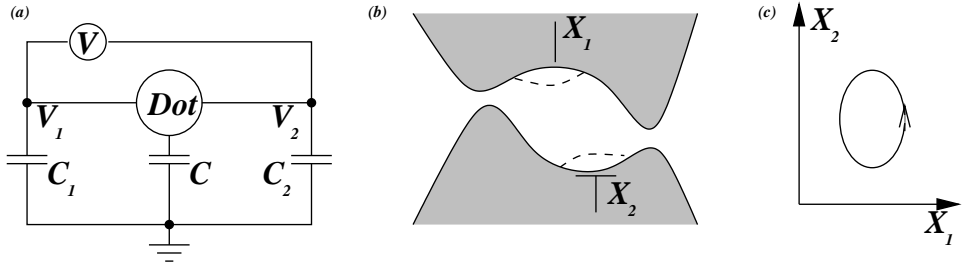} \caption{\label{fig:scheme}(a) Schematics of
the voltage measurement set-up: Quantum dot and reservoirs with
potentials $V_{1(2)}$ are capacitively coupled to the ground via $C$
and $C_{1(2)}$; (b) Weak time-dependent potentials $X_{1,2}(t)$
generate pumped dc current (voltage); (c) The pumped voltage $\bar
V$ is proportional to the area of the contour in parameter space of
the voltages $X_{1,2}(t)=X_{1,2}\cos(\o t+\phi_{1,2})$ (shown for
$\phi_1-\phi_2=\pi/2$).}
\end{figure}

Quantum pumping exploits the sensitivity of quantum interference to
variations of parameters of the scattering matrix. The variation of
two parameters $X_{1,2}(t)=X_{1,2}\cos(\o t+\phi_{1,2})$ oscillating
with the same frequency but out of phase generates a dc-current
\cite{spivak,pump,ZSA,SAA,avron,VAA,VA,moskalets}, if the oscillating
scatterer is in a zero-impedance external circuit, or a dc-voltage
\cite{Switkes}(if the pump operates in an external circuit with
finite impedance). Oscillating parameters $X_{1,2}$ are due to,
\emph{e.g.} voltages applied to the gates that form the shape of the
dot. Here we briefly discuss quantum pumping, see Fig.
\ref{fig:scheme}, when the dc voltage generated by the pump is
measured, similar to the experiment by Switkes \emph{et al.}
\cite{Switkes}. For slow variations of the shape of the dot we can use
the scattering matrix $\S(\e)$ and later include
Coulomb interactions self-consistently\cite{pump,parametric}. The
Coulomb potential is found from the condition of charge
conservation. Using a global gauge transformation it can be shifted
into the phases of the the scattering states. However, such a
transformation in itself does not pump electrons. The potential can
not be considered as an additional independent pump parameter
similar to $X(t)$.

We consider a two-terminal quantum dot with two weak potentials
$X_{1,2}$ varied with the same frequency $\o$. In order to derive a
formula for the dc voltage $V$, we use a simple model for the
quantum dot and the two electron reservoirs, see Fig.\
\ref{fig:scheme} a. The dot and the reservoirs $1$, $2$ are
connected to a screening gate via capacitances $C$ and $C_{1,2}$.
Following Refs.\ \onlinecite{buttiker2,Buettiker1}, we introduce the
emissivity $e\DD q(\a)/\DD X_j$  and $e\DD q(\a)/\DD E$, which is
the charge that exits the dot through a point contact $\a$
($\a=1,2$) when the parameter $X$ or chemical potential $E$ is
changed:
\begin{eqnarray}\label{eq:emissivity}
\frac{\partial q(\a)}{\partial X} = \frac{1}{2\pi i}\Tr
  \S^\dagger\L_\a\frac{\partial\S}{\DD X},\,\,
  \frac{\partial q(\a)}{\partial E} = \frac{1}{2\pi i}\Tr
  \S^\dagger\L_\a\frac{\partial\S}{\DD E}.
\end{eqnarray}
The total current $I_1$ flowing through contact $1$ is
\begin{eqnarray}
  C_1\frac{dV_1(t)}{dt}=I_1(t) &=& e\sum_{i=1}^2 \frac{\DD q(1)}{
\DD X_i}\frac{dX_i}{dt}+\frac{e^2}{h}g[V_1(t) - V_2(t)] ,
\end{eqnarray}
where $g=\Tr \S^\dagger\L_1\S\L_2$ is the conductance of the quantum
dot. A similar expression determines $I_2$. For slow variations of
$X$ we can restrict ourselves to first order time-derivatives only.
Since $I_j=C_jdV_j/dt$, in the limit $\omega \ll G/C_{1,2}$ we have
$I_1(t) = \eta I_2(t)$, where $\eta = C_{1}/C_{2}$ is a numerical
coefficient describing the capacitive division between the two
reservoirs. The dc voltage for small periodic parameters with
amplitudes $X_{1,2}$ and phase difference $\phi=\phi_1-\phi_2$ is
 $\bar V=(h\o/4e)X_1 X_2
\bar v \sin \phi$, where \cite{PB}
\begin{eqnarray}\label{eq:resVint}
\fl \bar v =\frac{1}{1+\eta} \left[\frac{\partial}{\partial
X_2}\left(\frac 1g \frac{\partial (q(1)-\eta q(2))}{\partial
X_1}\right)-\frac{\partial}{\partial X_1} \left(\frac
1g\frac{\partial (q(1)-\eta q(2))}{\partial X_2}\right)\right].
\end{eqnarray}

The derivatives with respect to $X_1$ and $X_2$ in Eq.
(\ref{eq:resVint}) are taken at constant values of the chemical
potential $E$ of the reservoirs. However, when the Coulomb
interaction is accounted for, it is the sum of chemical potential
and local self-consistent Hartree potential $U$, which is constant.
Therefore, in the presence of Coulomb interactions the expressions
for the transmitted charge should take this into account
\cite{parametric}. In the limit of weak intra-dot interactions,
$e^2/C\ll \Delta$, one still can apply the non-interacting theory
and take all derivatives with respect to shape-varying potentials
$X_{1,2}$ at fixed value of $E $. On the other hand, in the strong
electron-electron interaction limit, $e^2/C\gg\Delta$, which is
usually relevant in experiments, transport occurs on the background
of almost constant charge $q$ in the dot. Following Brouwer
\cite{pump} we can relate derivatives with respect to $X$ at
constant $E $ and $q$ and in this way include a Hartree potential of
arbitrary strength.

First we note that the applied potentials $X$, the chemical
potential $E $ and the charge of the dot $Q$ are not independent
variables. If we denote the derivative of some function $F$ with
respect to $X$ at constant $Y$ as $(\DD F/\DD X)_Y$, then
\begin{equation}
 \fl d F(E,Q,X) =dX\left(\frac{\DD F}{\DD X}\right)_E +\left(\frac{\DD F}{\DD E}\right)_X\left(dX\left(\frac{\DD E}{\DD X}\right)_q
 + d q\left(\frac{\DD q}{\DD E}\right)_X
\right),
\end{equation}
which allows us to write
\begin{eqnarray}\label{eq:derivation}
 \left(\frac{\DD}{\DD X}\right)_q  &=&\left(\frac{\DD}{\DD X}\right)_E  +\left(\frac{\DD E}{\DD X}\right)_q\DD_E
 =\left(\frac{\DD}{\DD X}\right)_E-\frac{(\DD q/\DD X)_E }{(\DD q/\DD E)_X} \DD_E .
\end{eqnarray}
\begin{figure}
\epsfxsize=0.55\hsize \hspace{0.25\hsize} \epsffile{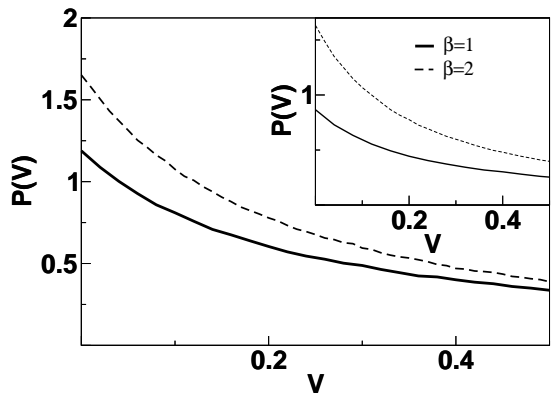}
\caption{\label{fig:Vplot} Mesoscopic distribution $P(\bar v)$ of
pumped voltage $\bar v$ for $N_1=N_2=1$ with and without
time-reversal symmetry (solid and dashed curves respectively). The
limits of strong, $C/e^2\ll \Delta$ (main figure, after Ref.
\cite{PB}), and weak interaction, $C/e^2\gg \Delta$ (inset), are
similar to each other.}
\end{figure}
The derivatives in the last equation in Eq. (\ref{eq:derivation})
are found as follows: the denominator is evaluated in the absence of
pumping (constant $X$) as a balance of particle current and
displacement current from the dot,
 $(\sum_{\a\b}G_{\a\b}(\o)+i\o C)dE = i\o edq_\o$, so that the derivative is
$\DD q_\o/\DD E =
(C/C_\m)\Tr{\cal N} $. The numerator is evaluated at constant $E$,
so the dependence of $\cal S$ on the pumping parameter $X$ is
crucial. Unlike the dc-pumped current, found by Brouwer to be
quadratic in pumped strengths \cite{pump}, the ac-component $I_\o$
is linear in $X$ and can be found by expansion of $\S([X(t)])$ in
$X$ to first order. As a result, $\DD q_\o/\DD X_\o=
 \Tr {\cal N}_X$. If analogously to Eqs. (\ref{ws},\ref{bu1}) we introduce  the
matrix ${\cal N}_X=(1/2\pi i)\S^\dagger\DD_X \S$ then
\begin{eqnarray}\label{eq:vgandc}
\frac{\partial E }{\partial X}&=& -\frac{C_\m}{C}\frac{\Tr{\cal
N}_X}{\Tr{\cal N}}.
\end{eqnarray}
 In the non-interacting limit $C\to\infty$
only the first term in the r.h.s. of Eq. (\ref{eq:derivation})
survives. In the opposite limit $C\to 0$ of a realistic quantum dot
the charging energy is large, $e^2/C\gg \Delta$. Then one finds that
the charge of an open dot essentially remains constant during the
pumping cycle, $I_1(t) = -I_2(t)$ for all time $t$. As a
consequence, the pumped voltage $\bar v$ loses its dependence on the
capacitive division $\eta$, the ratio $C_\m/C\to 1$ and therefore
the derivatives with respect to $X$ are
\begin{eqnarray}\label{eq:deriv}
(\DD_X )_q=(\DD_X )_E - \frac{\Tr{\cal N}_X}{\Tr{\cal N}}\DD_E.
\end{eqnarray}
Using Eqs. (\ref{eq:derivation},\ref{eq:vgandc}) and using the
emissivity $\DD q/\DD E$ (\ref{eq:emissivity}) we reformulate Eq.
(\ref{eq:resVint}) for arbitrary interaction. The distribution of
${\cal N}_X$ in the numerator of Eq. (\ref{eq:deriv}) is known
\cite{waves}, so that one can perform numerical integration for
few-channel dots or use the diagrammatic technique
\cite{PietBeenakker} to find the distribution of relevant
quantities, \emph{e.g.} the pumped voltage $\bar v$.

The pumped voltage $\bar v$ is zero when averaged over an ensemble
of dots, and the width of the distribution $P(\bar v)$ is expected
to diminish with growing $N$, similarly to the distribution $P(\bar
i)$ of the pumped current \cite{pump}. Ref. \onlinecite{PB} finds
that in the multi-channel limit the distribution has a Gaussian
shape with r.m.s. $\langle \bar v^2 \rangle^{1/2}=(2/\pi N^2)$,
since the fluctuations of conductance become small and the
conductance takes its classical value, $g=N/2$. However, in the
few-channel limit $N_1=N_2=1$ the mesoscopic distribution is wide,
see Fig. \ref{fig:Vplot}. The distributions in the limits of weak
and strong Coulomb interaction, treated with the help of  Eq.
(\ref{eq:deriv}), are similar to each other (in the weak interaction
limit we take symmetric capacitive division, $\eta=1$).
Particularly, for $\b=2$ numerics gives almost identical curves.
Thus the self-consistent internal potential does not change
distributions significantly, a conclusion that, for pumped currents,
was  reached  previously in Ref. \cite{pump}.


\section{Shot noise of photon excited electron-hole pairs}\label{sec:acnoise}
\begin{figure}
\epsfxsize=.55\hsize \hspace{0.25\hsize} \epsffile{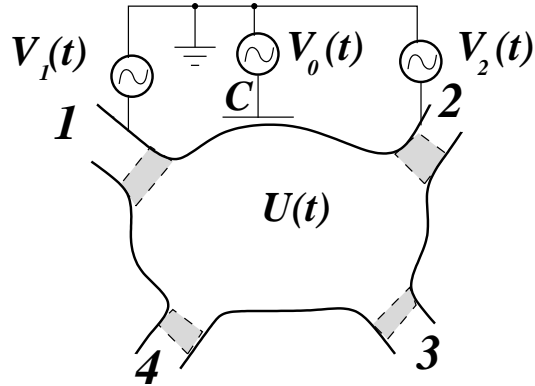}
\caption{\label{fig:4dot} Schematics of a noise measuring set-up. A
four-terminal ($M=4$) chaotic dot is subject to oscillating
potentials $V_\a(t)$ at contacts $\a=1,2$ and coupled to a gate with
a time-dependent potential $V_0(t)$, via a capacitance $C$. The
internal potential of the dot is $U(t)$. Tunneling elements (shaded
areas) partially close the dot.}
\end{figure}

In this section we consider the charge fluctuations and the
zero-frequency current noise generated by oscillating potentials
applied to the contacts and the gate of a multi-channel chaotic
sample with $M\geq 4$ leads (see Fig. \ref{fig:4dot}). Several
contacts are subject to periodic ac-potentials $V_\a(t)=V_\a\cos(\o
t+\phi_\a)$ at the same frequency $\o$ but possibly with different
phases $\phi_\a$. We consider the case $e V_\a \ll\hbar\o$, where
the amplitude of the applied potentials $e V_\a $ is small compared
to the modulation quantum $\hbar \omega$. The oscillating potentials
excite the electron gas in the contacts by creating electron-hole
pairs. These photon-excited carriers are then reflected at the
sample or transmitted through the sample into another contact. There
is no dc-current linear in voltage.

We find effect of Coulomb interactions on the noise through such a
multi-channel, ac-biased (partially) open quantum dot for arbitrary
frequencies $\o$ and interaction strength, see Fig. \ref{fig:4dot}.
Since in an experiment with dots or similar
structures \cite{glattli, dicarlo} frequencies $\o$ might be comparable to the
inverse dwell time, one can not restrict oneself to first energy
derivatives only. Consequently, the Wigner-Smith matrix is not as
useful for arbitrary $\o$, since we have to consider correlations of
electrons at quite different energies. However, in the limit $N\gg
1$ for open, or high total transmission $\Tr\hat \G\gg 1$ for
partially open cavities we can investigate correlations at arbitrary $\o$.

As emphasized in the Introduction, the Coulomb interaction is very
important for frequency-dependent transport. In the non-interacting
theory the noise of photon-assisted carriers was discussed by
Lesovik and Levitov\cite {LL} and Pedersen and B\"uttiker
\cite{pedersen} for energy-independent scattering matrices. This is
a good approximation if the frequency $\o$ is sufficiently small,
$\o\dwell\ll 1$. The sample-specific results of Refs. \cite{LL,
pedersen} are in a good agreement with an experiment by Reydellet
\emph{et al.} \cite{glattli} in quantum point contacts. However, if
an experiment is performed at high frequencies, such that
$\o\dwell\sim 1$, the non-interacting theory provides a  result
which is not gauge-invariant \cite{PSB}. In reality, the
frequency-dependence should be treated by taking Coulomb
interactions into account. In the set-up shown in Fig.
\ref{fig:4dot} we consider the current correlations in the leads
$\l,\m$ using scattering theory for chaotic systems
\cite{Beenakker,review}.

The rate of photon-generated pairs is small \cite{RPB}. If the pairs are
split into different contacts, they generate current noise.
For a single ac-voltage this noise corresponds to electron-hole
correlations of elements of a pair. For several ac-bias
voltages an adjustment of the phase shift between them allows one to
maximize (minimize) noise and extract information about the
scattering matrix \cite{RPB}. This is why it is interesting to find
current noise correlations $S_{\l\m}$ between $\l$-th and $\m$-th
lead for a multi-lead
(partially) open quantum dot. Similar results were obtained by
Samuelsson and the authors in Ref. \cite{PSB} for completely open
dots.

Below we first discuss shot noise for open quantum dots and show why
it is necessary to take the Coulomb interaction into account.
Subsequently we discuss the noise correlations to leading order in
$\Tr\hat\G$ for partially open dots, where the total transparency is
still large $\Tr\hat \G\gg 1$. In this limit the mesoscopic
fluctuations of the noise $S$ can be neglected, so that the result
gained from mesoscopic averaging is representative. Corrections to
the noise due to the symmetry appear in sub-leading orders, so we do
not distinguish here between $\b=1,2$. The coherent open dot with
$N$ channels is fully characterized by its  $N\times N$ scattering
matrix $\cal S$. Scattering is spin-independent and the results
given below are presented for a single spin direction. The dot is
assumed fully coherent and the effects of inelastic scattering and
dephasing are discussed in Ref. \cite{PSB}. We use units $e=h=k_{\rm
B}=1$.

We start by expressing the sample-specific noise $S_{\l\m}$ in terms
of energy-dependent scattering matrices ${\cal S}(\e)$, amplitudes
of applied voltages $V_\a$, their phases $\phi_\a$ and the electron
distributions in the leads $f_\a(\e)$. Ref. \cite{RPB} finds,
\begin{eqnarray}\label{eq:ij}
\fl S_{\lambda\mu}&=&\frac{e^2}{2\o}\mbox{Re }\sum_{m,\,\alpha\beta}
\Tr\left({\bf A}_{\alpha\beta}(\lambda,m\o){\bf
A}_{\beta\alpha}(\mu,0)\right)\left(\d_{m,0}V_\a^2-V_\a
V_\b\d_{m,1}e^{i(\phi_\b-\phi_\a)}\right),
\end{eqnarray}
where ${\bf A}(\l,\e)=\L_\l-\S^\dag(\e) \L_\l \S(\e)$. In the
non-interacting low-frequency limit, such that $\S$ is
energy-independent, the ensemble averaged noise is \cite{PSB}
\begin{eqnarray}\label{eq:Main}
 S_{\l\m}^{\rm o}&=& \frac{2{\bar
G}_{\lambda\mu}}{N^2\o}\left( N\Tr \ V^2-|\Tr V e^{i\phi}|^2\right),
\end{eqnarray}
where we introduced the averaged conductance of the open quantum dot
${\bar G}_{\l\m}=(N_\l\d_{\l\m}-N_\l N_\m/N)$ and  diagonal matrices
of the amplitudes $V={\rm diag}(V_1\L_1,...,V_M\L_M)$ and phase
shifts $\phi={\rm diag }(\phi_1\L_1,...,\phi_M\L_M)$. This result is
gauge-invariant, that is an arbitrary uniform shift of all Fourier
components of potentials $V_{\a,\o}=V_\a \exp(i\phi_\a)$ by some
(complex) potential $U_0$ does not have any physical consequence. We
notice here that although formally the 2nd term in Eq.
(\ref{eq:Main}) is of order ${\cal O}(1/N^2)$, the traces in the
numerator can be of order $N$, so that both terms may be of the same
order. For example, when a global shift of potentials is applied
both terms are equally important.

However, if we are at sufficiently high frequencies, $\o\dwell\sim
1$, the energy-dependence of $\S$ is important, and the
noise spectra are given by
\begin{eqnarray}\label{eq:noninvar}
S_{\l\m}&=& S_{\l\m}^{\rm o}+\frac{2{\bar
G}_{\lambda\mu}}{N^2\o}\frac {|\Tr V
e^{i\phi}|^2}{1+(\o\dwell)^{-2}} .
\end{eqnarray}
These spectra are not gauge-invariant because of the 2nd term. This
is a consequence of the fact that we neglected the internal
potential $U$ of the dot and the effect of external gates with
potential $V_0\cos(\o t+\phi_0)$ capacitively coupled to the dot.
Therefore now we consider explicitly the effect of the induced
internal potential $U_\o$. The sample-specific formula (\ref{eq:ij})
for the non-interacting system can be used again if we shift all
potentials $V_\o\to V_\o-U_\o$, including unbiased leads.

To proceed we find the \emph{mesoscopically averaged} potential
$\bar U_\o$ for a partially open dot. The mesoscopic fluctuations of the
potential (\ref{eq:UU}) around the average are small due to the
large parameter $\Tr\hat\G\gg 1$, so we are justified to use $\bar
U_\o$ if we are interested in current correlations only to leading
order. As a consequence, any effects related to time-reversal
symmetry are neglected both in $\bar U_\o$ and, in the end, in the
noise $S_{\l\m}$. At high frequencies, $\o\dwell\sim 1$, we have to
fully account for the frequency dependence of $\bar U_\o$. The
leading order of Eq. (\ref{eq:double}) for the pair correlator of
energy-dependent scattering matrices \cite{ABG, iop} allows us to
find the average of the potential $U_\o$ and its derivatives,
\begin{eqnarray}\label{eq:Uomega}
\fl{\bar U}_\o&=& V_0 e^{i\phi_0}+\frac{C_\mu/C }{1-i\o R
C_\mu}\left(\frac{\Tr\hat\G V e^{i\phi}}{\Tr\hat\G}-V_0
e^{i\phi_0}\right),\,\,\, \frac{\DD \bar U_\o}{\DD
V_{\a,\o}}=\frac{C_\mu/C }{1-i\o R
C_\mu}\frac{\Tr\hat\G_\a}{\Tr\hat\G},
\end{eqnarray}
where we introduced $R = 1/\Tr \G$, which corresponds to the
ensemble-averaged charge relaxation resistance $\langle R_q\rangle $
given in  Eq. (\ref{rq}) for a partially open dot, see also Eq.
(\ref{bee}). For an open dot, $\hat \G\equiv \L$ the result
corresponds to that of \cite{PSB}. We see that the dwell time
$\dwell=1/(\Tr\hat \G\Delta)$ of matrix correlators is replaced by
the $RC$-time of charge relaxation inside the dot, $\tau=\langle
C_\m \rangle R$. This substitution usually occurs in leading order
in $N$. The $\o\to 0,\G=1$ limit of Eq. (\ref{eq:Uomega}) could be
easily obtained from Eq. (\ref{eq:U}) using the Wigner-Smith matrix
and ensemble averaging (using results of Ref. \cite{waves}). From
Eq. (\ref{eq:Uomega}) for $\G=1$ and a shift of all potentials
$V_{\o,\a}$ as described above, we obtain with a little algebra
\cite{PSB}:
\begin{eqnarray}\label{eq:Sint} S_{\l\m}&=&\frac{2{\bar G}_{\lambda\mu}}{N^2\o}\left( N\Tr \
V^2-|\Tr V e^{i\phi}|^2+\frac {|\Tr V e^{i\phi}-N V_0
e^{i\phi_0}|^2}{1+(\o\tau )^{-2}}\right).
\end{eqnarray}
The second term in this equation is obviously gauge invariant and
should be contrasted with that in Eq. (\ref{eq:noninvar}).
Experimentally, one usually has $\o\tau\ll 1$ because the Coulomb
interaction is sufficiently strong, and the 2nd term in Eq.
(\ref{eq:Sint})
 vanishes. Thus the non-interacting low-frequency limit
(\ref{eq:Main}) is recovered. The fact that the limits of strong
interaction at $\o\tau\ll 1$ and weak interaction at $\o\dwell\ll 1$
provide the same result explains why the results of the
non-interacting theories of Refs. \cite{LL,pedersen} are in good
agreement with experiment \cite{glattli}.
At sufficiently high frequencies, which are in the range of
modern experiments, the difference will become
apparent.

If one now considers a partially open quantum dot with channel
transparencies $\G_i\neq 1$, a similar treatment is possible. For
simplicity we consider the case when all channels have the same
transparency $\G_i=\G$. One has to use Eq. (\ref{eq:quadruple}) for
the correlators of scattering matrices. As above, the external gates
with potential $V_0(t)=V_0\cos(\o t+\phi_0)$ are capacitively
coupled to the dot via  a capacitance $C$. To demonstrate the
gauge-invariance of the final result for the noise, we assume that
the current-measuring leads $\l,\m$ are also biased. Then the
non-interacting noise is given by an expression analogous to Eq.
(\ref{eq:Main}):
\begin{eqnarray}\label{eq:S_tunnel}
\fl S_{\l\m}^{\rm
o}(\G)=\frac{2\G}{N^2\o}\left(N_\l\d_{\l\m}-\frac{\G N_\l
N_\m}{N}\right)\left(N\Tr \ V^2-|\Tr\ V
e^{i\phi}|^2\right)+\frac{\G(1-\G)}{N^2\o}\nonumber
\\  \mbox{} \times\left(N_\l\d_{\l\m}-\frac{
2N_\l N_\m}{N}\right)|\Tr V e^{i\phi}-NV_\l e^{i\phi_\l}|^2+\left.
...\right|_{\l\leftrightarrow\mu},
\end{eqnarray}
where the last term (...) is obtained from the 2nd by swapping
$\l\leftrightarrow\mu$. If we now take into account the frequency
dependence of the scattering matrices similarly to the completely
open dot, and use Eq. (\ref{eq:quadruple}) we find:
\begin{eqnarray}\label{eq:Sint_tunnel}
\fl S_{\l\m}(\G)=\frac{2\G}{N^2\o}\left(N_\l\d_{\l\m}-\frac{\G N_\l
N_\m}{N}\right)\left(N\Tr \ V^2-|\Tr\ V e^{i\phi}|^2+ \frac{|\Tr V
e^{i\phi}-NV_0 e^{i\phi_0}|^2}{1+(\o\tau )^{-2}}\right)\nonumber \\
\fl \mbox{} +\frac{\G(1-\G)}{N^2\o}\left(N_\l\d_{\l\m}-\frac{2 N_\l
N_\m}{N}\right) \left|\frac{\Tr V e^{i\phi}- i\o\tau NV_0
e^{i\phi_0}}{1-i\o\tau}-N V_\l e^{i\phi_\l}\right|^2\left.+
...\right|_{\l\leftrightarrow\mu}.
\end{eqnarray}
As an application, consider the cross-correlations in the unbiased
leads, $\l\neq \m$ and $V_{\l}=V_\m=0$, when $V_0=0$:
\begin{eqnarray}\label{eq:Scross_tunnel}
S_{\l\m}(\G)=-\frac{2\G}{N^2\o}\frac{N_\l N_\m}{N}\left(N\G\Tr \
V^2+\frac{2-3\G}{1+\o^2\tau^2}|\Tr V e^{i\phi}|^2\right).
\end{eqnarray}
Notice that the interactions can not change the negative sign of
cross-correlations. For sufficiently transparent barriers, $\G>2/3$,
the Coulomb interactions diminish the absolute value of the
cross-correlation (46). This is easy to see from the dependence of
the RC-time on the geometrical capacitance, $\tau^{-1}=\Tr\hat
\G(\Delta+e^2/C)$. Indeed, the interactions in the form of the
Hartree potential lead to additional (displacement) currents, so
that their fluctuations "damp" particle current fluctuations [79].
On the contrary, for $\G<2/3$ the 2nd term in Eq. (46) becomes
positive and Coulomb interactions \emph{enhance} the noise. Indeed,
for $\G\to 0$ the response of the dot becomes capacitive rather then
resistive (see the denominator in Eq. (42)), and the displacement
currents fluctuate out of phase.

Next, similarly to van Langen and B\"uttiker \cite{VLB}, we consider
the exchange correlation $P^{\rm ex}(\phi)$. This is the difference
of the noise $S_{34}$ with bias $V_{1,\o}=V_{2,\o}^*=Ve^{i\phi/2}$
and of the sum of the noises for $V_1=0,V_2=V$ and $V_1=V,V_2=0$.
For a four-channel dot we find
\begin{eqnarray}\label{eq:Pex}
\fl P^{\rm ex}(\phi)\equiv
S_{34}(Ve^{i\phi/2},Ve^{-i\phi/2})-S_{34}(V,0)-S_{34}(0,V)=\frac{V^2\cos\phi}{16\o}\frac{\G(3\G-2)}{1+\o^2\tau^2}.
\end{eqnarray}
For a dc-biased dot, the inversion of the sign of the exchange
correlation at $\G=2/3$ was found previously \cite{VLB}. Notice that
Coulomb interactions enhance the absolute value of the exchange
correlation (\ref{eq:Pex}).


\section{Discussion of related works}\label{sec:discussion}

In this section we briefly allude to alternative approaches to treat
the Coulomb interaction and discuss the applicability of the results
presented above. In this paper we used the RPA approach to the
Coulomb interaction which is built into the scattering matrix
approach by B\"uttiker and co-workers
\cite{buttiker2,Buettiker1,BC,BJLTP}. Another, often favoured path
is the Hamiltonian approach. For non-interacting electrons it is
completely equivalent to the scattering approach
\cite{LW,Brouwer1995} and which method one uses is a matter of
taste. Aleiner, Brouwer, and Glazman \cite{ABG} demonstrated how the
Coulomb interactions can be considered in the Hamiltonian approach.
The interaction Hamiltonian has a hierarchical structure: its
universal part depends only on the number of electrons inside and
the energy-scale $e^2/C$. The non-universal terms due to
non-uniformity of the potential in the dot \cite{ABG,BMM} are small,
$\ll \Delta$. Writing the Hamiltonian of the total system as a sum
of Hamiltonians of the closed interacting dot, of the leads, and the
coupling term, one can express transport quantities and proceed with
a Green function formulation\cite{ABG,BLF}. Then the Coulomb
potential is usually considered in several theoretical limits:
either for arbitrary number of channels $N$, if the interaction
strength is small
\cite{Flensberg1993,Matveev,Furusaki_Matveev,Furusaki1995}, or in a
diagrammatic expansion in $1/N\ll 1$ for arbitrary interaction
strength \cite{Brouwer1999,Yeyati,GZ,GZ_2004}. One can see that the
total transmission $\Tr\hat \G$ is an essential parameter of the
problem.

It is known from the literature that the mean-field treatment, which
renormalizes the interaction line, can not lead to such effects as,
\emph{e.g.} charge quantization in blockaded quantum dots \cite{AA}.
However, in the multi-channel limit $N\gg 1$ the exchange terms can
be neglected, and the Hartree potential of the dot provides the
leading effect of the Coulomb interaction. Therefore, one expects
that in a crossover from multi-channel systems to the Coulomb
blockade regime, the exchange contributions may become important.
Only very recently did Brouwer, Lamacraft and Flensberg \cite{BLF}
consider a unified approach using a Keldysh technique and scattering
matrix theory. They demonstrated that the exchange diagrams may
become important when the number of conducting channels $N$ becomes
small, and so for the dot with poor transmission a Hartree treatment
is not enough. For the variance of the electro-chemical capacitance
$C_\m$ and the density of states $\nu(\e)$, as well as for the
variance of the charge pumped by a quantum pump, the corrections due
to exchange were found to be ${\cal O}(1/N)$ for $N\gg 1$. The
self-consistent theory by B\"uttiker and co-workers is reproduced
for open coherent dots to leading order in $1/N$. Inclusion of
exchange terms to arbitrary order is a challenging task, but the
authors \cite{BLF} argue that the expansion in $1/N$ is still useful
even for two-channel systems ($N=4$ if spin degeneracy is taken into
account) and the proposed diagrammatic technique in $1/N$ can still
be used for few-channel dots. As another example of an application
of a $1/N$-expansion to few-channel open dots, we mention recent
work by Vavilov, DiCarlo and Marcus \cite{Vavilov}. There the strong
Coulomb interaction was treated as a self-averaging Hartree
potential and was transformed into the phases of scattering states
in the leads. Good agreement with experiment is reported for the
variance of the photovoltaic current, if the leading in $N$ term is
used when $N=2$.

To summarize, we expect that the self-consistent treatment of the
Coulomb interaction within the scattering approach is still
qualitatively correct down to $N=1-2$ channels. Measurement of
transport involving sufficiently high frequency $\o$ would clarify
the question of the importance of the exchange terms
\cite{gabelli,dicarlonoise}.
\section{Conclusions}\label{sec:conclusions}
In this work we have focused on transport problems in which dynamic
charge fluctuations play an essential role. Chaotic quantum dots
provide an example of a generic conductor which serves to illustrate
the basic role which the long range Coulomb interaction plays in
problems of this type. The isotropic chaotic scattering allows us in
many cases to treat interaction effects with just a single
potential. As a consequence we can simplify the theoretical
discussion to an extent which would be unrealistic and inappropriate
for less generic conductors. We have pointed out that at low
frequencies the charge fluctuations can be described with the help
of a generalized Wigner-Smith matrix in which we replace the energy
derivative with a derivative with respect to the local electric
potential. Together with a self-consistent treatment of the charge
response this leads to a charge operator, which determines
capacitances, kinetic inductances, the weakly non-linear current
voltage characteristics, charge relaxation resistances and the low
frequency noise power of charge fluctuations. We have shown how this
approach is applied to specific examples, like the rate of
relaxation and decoherence of a charge qubit near a chaotic cavity,
or quantum pumping. We have illustrated how the approach is applied
to transport problems in which the frequency is not small,
discussing the shot noise of photon-excited electron-hole pairs. The
unified description of these problems demonstrates the generality of
the approach, and makes it a useful tool for future investigations.
On the experimental side there is an increasing interest to explore
dynamic phase-coherent transport. This interest derives not only
from the fact that the dynamics of small structures is largely an
unexplored area, but also from the desire to use quantum coherent
structures to perform useful tasks, like quantum information
processing, and to perform these tasks as fast as possible.

\section*{Acknowledgement}
We acknowledge stimulating discussions with Piet Brouwer, Leo
DiCarlo, Julien Gabelli, Christian Glattli, Michael Moskalets, and
Eugene Sukhorukov. This work was supported by the Swiss National
Science Foundation and the Marie Curie RTN on Nanoscale Dynamics and
Quantum Coherence.

\appendix
\section{Matrix correlators for energy-dependent matrices with non-ideal
leads}\label{sec:appendix}

 In this Appendix we present several
correlators of the scattering matrix $\S$ of chaotic partially open
quantum dot, when the contacts have transmission $\G_i\leq 1,
i=1,\ldots,N$. These correlators are used to find conductance
auto-correlations and noise through a biased dot. The imperfect
transmission of the contacts is characterized by an $N \times N$
reflection matrix $\bar {\cal S}$ for which we take the simple form
\begin{equation}
  \bar {\cal S}=\langle{\cal S}\rangle=(1-\hat\Gamma)^{1/2},
  \label{eq:rc}
\end{equation}
where $\hat\Gamma$ is an $N \times N$ diagonal matrix containing the
transmission coefficients $\Gamma_j\in[0,1]$. In order to describe
coherent energy-dependent scattering with nonideal leads, we use the
following method. An $N\times N$ matrix ${\cal U}(\e)$ of a fully
open quantum dot is found using stub model
\cite{waves,PietMarkus,BB_96}. Energy-independent reflection from
the contacts is modeled by $\bar {\cal S}$, but since the carrier
can enter the dot with probability $\G^{1/2}$, its scattering is
described by \cite{Brouwer1995,FriedmanMello}
\begin{eqnarray}
  {\cal S}(\e) &\equiv&{\bar{\cal S}}-\d{\cal S}(\e)=\bar {\cal S}-
  \Gamma^{1/2}{\cal U}(\e)\left(1-(1-\G)^{1/2}{\cal U}(\e)\right)^{-1}
  \Gamma^{1/2}.
  \label{eq:rcstub}
\end{eqnarray}
The matrix correlators below are presented for the case of fully
broken time-reversal symmetry, $\b=2$.  When this symmetry is
present, one should add terms which correspond to permutations of
indices, so that \emph{e.g.} for 2-matrix correlator, a Cooperon
contribution is added, and for the irreducible four-matrix
correlator (Hikami box) each term containing four $\delta$-functions
should be supplemented by 7 other terms, obtained by permutations of
3 pairs of indices.

\begin{figure}
\epsfxsize=1.0\hsize \epsffile{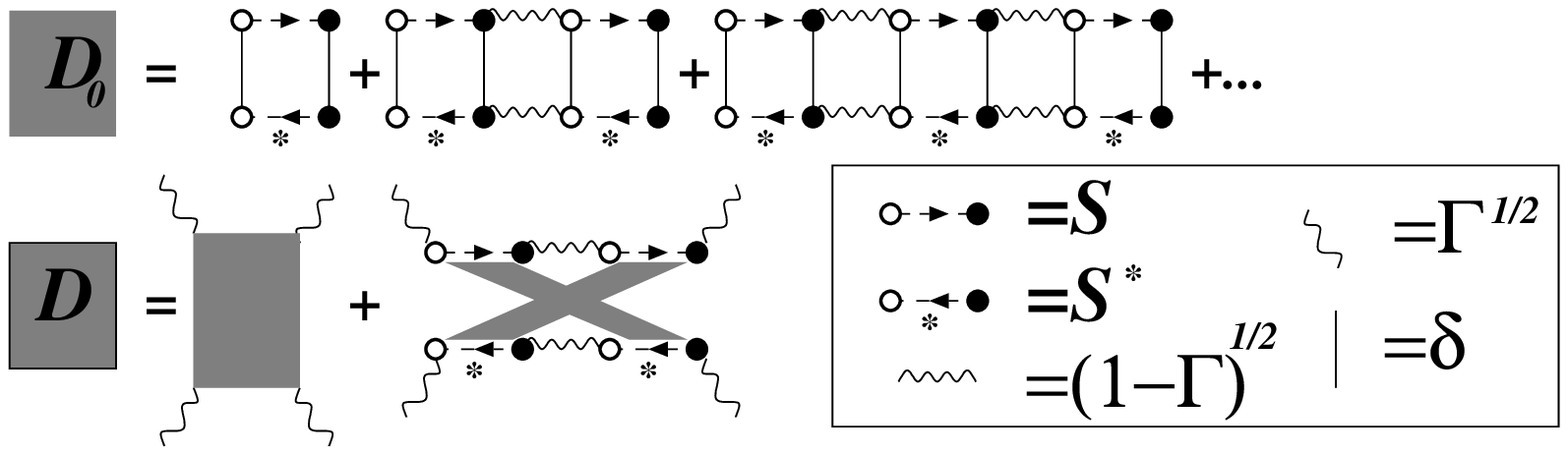}
\caption{\label{fig:diffuson} Top: Ladder diagrams for leading order
$\Dif_0$ contribution to the full 2-matrix correlator
$\Dif=\langle(\d\S)(\d\S^*)\rangle$. Bottom: Leading and sub-leading
terms for $\Dif$. The 2nd term contributes, \emph{e.g.}, to var
$G$.}
\end{figure}

\begin{figure}
\epsfxsize=1.0\hsize
\epsffile{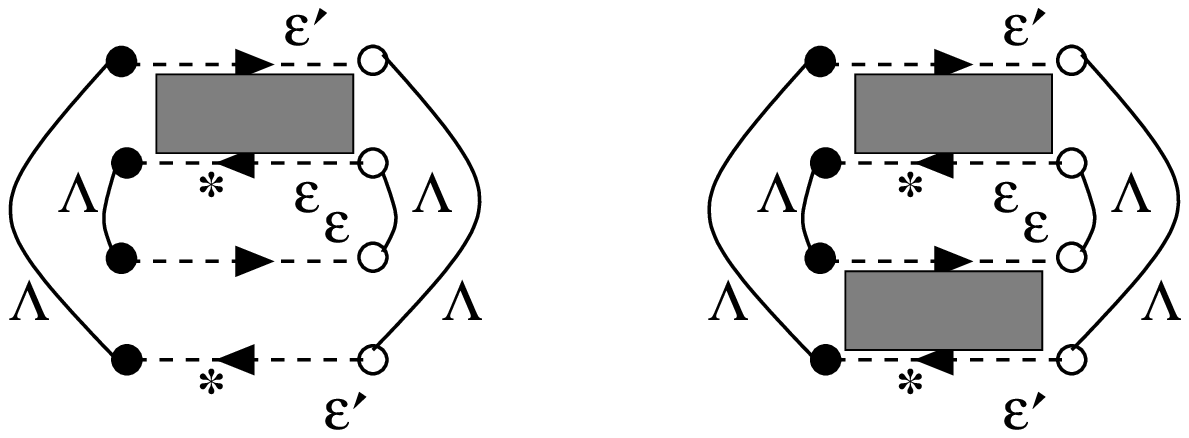} \caption{\label{fig:conductance} Diagrams
for the evaluation of $\langle\langle G(\e)G(\e')\rangle\rangle$:
(left) A single diffuson $\Dif$ connects one pair of matrices, the
others being averaged to $\langle\S(\S^*)\rangle$; (right) $\Dif^2$
gives diagrams of the same order. For $\b=1$ an additional
contribution comes from the substitution $\Dif\to\Coop$ in these
diagrams.}
\end{figure}

The basic element of diagrammatic technique is a 2-matrix
correlator, an ensemble-averaged product of two scattering matrix
elements \cite{iop}. For convenience we introduce a shorthand
notation $ D (\e,\e')=1/(\Tr\hat\G-i(\e-\e'))$, where $\e,\e'$ are
normalized by $2\pi/\Delta$. The correlator $\langle \d{\cal
S}\d{\cal S}^*\rangle$, shown on Fig. \ref{fig:diffuson}, is
\begin{eqnarray}\label{eq:dSdS}
\fl&&\langle(\d{\cal S})_{ij}(\e)(\d{\cal S})_{kl}^*(\e')\rangle
=\G_i\G_l D (\e,\e')\left(\d_{ik}\d_{jl}
+\sqrt{(1-\G_i)(1-\G_l)}D (\e,\e')\d_{ij}\d_{kl}\right).
\end{eqnarray}
Notice that the 2nd term is actually sub-leading in $\Tr\hat\G$, but
we keep it for the reasons to be explained later. Using Eq.
(\ref{eq:dSdS}) we find
\begin{eqnarray}\label{eq:double}
\langle{\cal S}_{ij}(\e){\cal S}_{kl}^*(\e')\rangle &=&
\sqrt{(1-\G_i)(1-\G_l)}\d_{ij}\d_{kl}+ \G_i\G_l D
(\e,\e')\d_{ik}\d_{jl}\nonumber \\
&&+\sqrt{(1-\G_i)(1-\G_l)}\G_i\G_l D ^2(\e,\e')\d_{ij}\d_{kl}.
\end{eqnarray}
The leading order term ( first line) in Eq. (\ref{eq:double}) was
previously found in Refs. \cite{ABG, iop}. The expressions for
correlators of $n$ matrix elements for $n>2$ are more complicated,
and below we present a 3-matrix correlator of
 $\d\S\d\S^*\d\S^*$ to leading order:
\begin{eqnarray}\label{eq:triple}
\fl &&\langle(\d{\cal S})_{ij}(\e)(\d{\cal
S})_{k_1l_1}^*(\e'_1)(\d{\cal S})_{k_2l_2}^*(\e'_2)\rangle =
\G_{i}\G_{j}D (\e,\e_1')D (\e,\e_2')\nonumber
\\ &&\mbox{}\times\left(\G_j\sqrt{1-\G_j}\d_{il_2}\d_{k_2l_1}\d_{k_1j}+\G_i\sqrt{1-\G_{i}}\d_{ik_2}\d_{l_2k_1}\d_{l_1j}\right).
\end{eqnarray}
 The 4-matrix correlator is equal to the reducible (disconnected)
part and the Hikami box denoted$\blacksquare $ for the irreducible
4-th moment:
\begin{eqnarray}\label{eq:quadruple}
\fl \langle(\d{\cal S})_{i_1j_1}(\e_1)(\d{\cal
S})_{i_2j_2}(\e_2)(\d{\cal S})_{k_1l_1}^*(\e'_1)(\d{\cal
S})_{k_2l_2}^*(\e'_2)\rangle =\blacksquare
+\G_{i_1}\G_{i_2}\G_{l_1}\G_{l_2}\nonumber \\
\fl  \mbox{}\times\left(D (\e_1,\e_1')D
(\e_2,\e_2')\d_{i_1k_1}\d_{j_1l_1}\d_{i_2k_2}\d_{j_2l_2}
+D (\e_1,\e_2')D
(\e_2,\e_1')\d_{i_1k_2}\d_{j_1l_2}\d_{i_2k_1}\d_{j_2l_1}\right),
\end{eqnarray}
\begin{eqnarray}\label{eq:Hikami}
\fl \blacksquare = \G_{i_1}\G_{i_2}\G_{l_1}\G_{l_2}D (\e_1,\e_1')D
(\e_2,\e_2')D (\e_1,\e_2')D (\e_2,\e_1') \left\{\delta_{i_1k_1}
\delta_{j_1 l_2} \delta_{i_2 k_2} \delta_{j_2 l_1} \right.\nonumber
\\ \fl \left. \mbox{}
\times\left([\Tr(\G^2-2\G)+i(\e_1-\e'_1+\e_2-\e'_2)]+(1-\G_{i_1})D
^{-1}(\e_1,\e_1')\right.\right.\nonumber
 \\ \fl \left. \left.\mbox{}+(1-\G_{l_2})D
^{-1}(\e_1,\e_2')+(1-\G_{l_1})D ^{-1}(\e_2,\e_1')+(1-\G_{i_2})D
^{-1}(\e_2,\e_2') \right) \right.\nonumber\\ \fl \left.\mbox{}
+\delta_{i_1k_2} \delta_{j_1 l_1} \delta_{i_2 k_1} \delta_{j_2
l_2}\left([\Tr(\G^2-2\G)+i(\e_1-\e'_1+\e_2-\e'_2)]+(1-\G_{l_1})D
^{-1}(\e_1,\e_1')\right. \right.\nonumber \\ \fl \left.\left.
\mbox{} +(1-\G_{i_1})D ^{-1}(\e_1,\e_2') +(1-\G_{i_2})D
^{-1}(\e_2,\e_1')+(1-\G_{l_2})D
^{-1}(\e_2,\e_2')\right)\right.\nonumber \\
\fl \left.\mbox{} +\left(\delta_{i_1k_1} \delta_{l_1 k_2}
\delta_{l_2 j_2} \delta_{i_2
j_1}(1-\G)_{i_2}^{1/2}(1-\G)_{l_1}^{1/2}+\delta_{i_1j_2} \delta_{i_2
k_2} \delta_{l_2 k_1} \delta_{l_1
j_1}(1-\G)_{i_1}^{1/2}(1-\G)_{l_2}^{1/2}\right)\right.\nonumber \\
\fl \left.\times
 \left[D ^{-1}(\e_1,\e_1')+D
^{-1}(\e_2,\e_2')\right] +\left(\delta_{i_1k_2} \delta_{l_2 k_1}
\delta_{l_1 j_2} \delta_{i_2
j_1}(1-\G)_{i_1}^{1/2}(1-\G)_{l_1}^{1/2} \right. \right.\nonumber
\\ \fl \left. \left.\mbox{}+\delta_{i_1j_2} \delta_{i_2 k_1}
\delta_{l_1 k_2} \delta_{l_2
j_1}(1-\G)_{i_2}^{1/2}(1-\G)_{l_2}^{1/2}\right)\left[D
^{-1}(\e_1,\e_2')+D ^{-1}(\e_2,\e_1')\right]\right\}.
\end{eqnarray}
We point out that the 4-matrix correlator does not simply reduce $N$
to $\Tr \hat \G$ or $\Tr \hat\G^2$, when the dot becomes partially
open. As a result, Hikami box found in Appendix A of Ref. \cite{iop}
for partially open dots reproduces Eq. (\ref{eq:Hikami}) only for
$\G=1$. One can use the correlators
(\ref{eq:double}--\ref{eq:Hikami}) to find, \emph{e.g.} the
two-terminal dimensionless conductance $G$ through a quantum dot and
its correlation function $\langle\langle
G(\e)G(\e')\rangle\rangle=\langle G(\e)G(\e')\rangle-\langle
G(\e)\rangle\langle G(\e')\rangle$ at different energies. To do this
we express conductance in terms of a traceless matrix $\Lambda$ as
\begin{eqnarray}
\fl G_{12}(\e)=\Tr
(\S^\dagger(\e)\L_1\S(\e)\L_2)=\frac{N_1N_2}{N}-\Tr
(\S^\dagger(\e)\Lambda\S(\e)\Lambda),\,\Lambda\equiv\frac{N_2}{N}\L_1-\frac{N_1}{N}\L_2.
\end{eqnarray}
Averaging, using 2-matrix correlators, yields the energy-independent
conductance:
\begin{eqnarray}
\langle G(\e)\rangle=\Tr
\hat\G\Lambda^2-\frac{\Tr^2\hat\G\Lambda}{\Tr
\hat\G}-\frac{\Tr\hat\G^2\Lambda^2}{\Tr\hat\G}\d_{\b,1}.
\end{eqnarray}
To find $\langle\langle G(\e)G(\e')\rangle\rangle$ we have to
calculate 2 diagrams shown in Fig. \ref{fig:conductance}.
\begin{eqnarray}\label{eq:condvar}
\fl\langle\langle G(\e)G(\e')\rangle\rangle
=\frac{2}{\b}\left(2\frac{\Tr\hat \G\,\Tr
\Lambda^4(1-\hat\G)\hat\G^2+\Tr^2
\Lambda^2(1-\hat\G)\hat\G}{\Tr^2\hat\G+(\e-\e')^2}+\frac{\Tr^2\Lambda^2\hat\G^2}{\Tr^2\hat\G+(\e-\e')^2}\right)
.
\end{eqnarray}
Notice that for the left diagram on Fig. \ref{fig:conductance} we
have to use a sub-leading term from Eq. (\ref{eq:dSdS}), which gives
the $\Tr^2 \Lambda^2(1-\hat\G)\hat\G$ term in Eq.
(\ref{eq:condvar}). The doubling of the left diagrams comes from
possibility to have correlators on the top and bottom. Due to the
convenient representation of $G$ in terms of traceless matrix
$\Lambda$, the irreducible averages (\ref{eq:triple}) and
(\ref{eq:Hikami}) do not contribute to the correlations. At
$\hat\G=\G\L, N_1=N_2=N/2$ and $\e=\e'$ we reproduce the result
$\mbox{var } G=(2-2\G+\G^2)/(8\b)$ obtained by Brouwer and Beenakker
\cite{PietBeenakker}, and at $\G=1, \e\neq\e'$  we have $\mbox{var }
G=(1/8\b)/(1+(\e-\e')^2/N^2)$ \cite{iop,Efetov95, Frahm95}.

To obtain Eqs. (\ref{eq:Sint}--\ref{eq:Sint_tunnel}) we need a
correlator (\ref{eq:quadruple}) combined in the form of ${\bf
A}_{\a\b}(\l,\e){\bf A}_{\b\a}(\m,\e')$, the matrices $\bf A$ are
defined after Eq. (\ref{eq:ij}).
 One can also use Eq.
(\ref{eq:quadruple}) in the limit $\o\to 0$ to obtain the noise in a
dc-biased dot \cite{VLB}.

\end{document}